\RequirePackage{rotating}
\documentclass[preprint]{emulateapj}
\usepackage{graphicx,times}
\usepackage{natbib}
\usepackage{amssymb,amsmath}
\usepackage{float}
\bibpunct{(}{)}{;}{a}{}{,}

\usepackage[T1]{fontenc}
\usepackage{pdftexcmds}

\def\hi{H{\textsc i} }

\def\co{$^{12}$CO}
\def\co13{$^{13}$CO}
\def\co18{C$^{18}$O}
\def\cm3{cm$^{-3}$}
\def\cm2{cm$^{-2}$}

\def\cm2{cm$^{-2}$}

\def\kms{km s$^{-1}$}

\def\nh3{NH$_3$}
\def\n2h{N$_2$H$^+$}
\def\hc3n{HC$_3$N}
\def\h2{H$_2$}
\def\nh{n(H$_2$)}

\usepackage{color}
\usepackage{cancel,soul,ulem}
\usepackage{rotating,multirow,amsmath}

\usepackage{enumerate}   
 \usepackage[flushleft]{threeparttable}  

\shorttitle{Pilot \hi Survey of Planck Galactic Cold Clumps with FAST }
\shortauthors{Tang et al.}

\begin{document}

\title{Pilot \hi Survey of Planck Galactic Cold Clumps with FAST }

\author{Ningyu Tang\altaffilmark{1,2}, 
Di Li\altaffilmark{1,2,3}, 
Pei Zuo\altaffilmark{4}, 
Lei Qian\altaffilmark{1,2},
Tie Liu\altaffilmark{5,6}, 
Yuefang Wu\altaffilmark{7},
Marko Kr$\rm\check{c}$o\altaffilmark{1,2},
Mengting Liu\altaffilmark{1,2},
Youling Yue\altaffilmark{1,2},
Yan Zhu\altaffilmark{1,2},
Hongfei Liu\altaffilmark{1,2},
Dongjun Yu\altaffilmark{1,2},
Jinghai Sun\altaffilmark{1,2},
Peng Jiang\altaffilmark{1,2},
Gaofeng Pan\altaffilmark{1,2},
Hui Li\altaffilmark{1,2},
Hengqian Gan\altaffilmark{1,2},
Rui Yao\altaffilmark{1,2},
Shu Liu\altaffilmark{1,2}, 
and FAST Collaboration
} 


\altaffiltext{1}{National Astronomical Observatories, CAS, Beijing 100012, China; {\it nytang@nao.cas.cn, dili@nao.cas.cn, lqian@nao.cas.cn}} 
\altaffiltext{2}{CAS Key Laboratory of FAST, National Astronomical Observatories, Chinese Academy of Sciences, Beijing 100101, China} 
\altaffiltext{3}{School of Astronomy and Space Science, University of Chinese Academy of Sciences, Beijing 100049, China}
\altaffiltext{4}{Kavli Institute for Astronomy and Astrophysics, Peking University, 100871 Beijing, PR China}
\altaffiltext{5}{Shanghai Astronomical Observatory, Chinese Academy of Sciences, 80 Nandan Road, Shanghai 200030, People’s Republic of China}
\altaffiltext{6}{East Asian Observatory, 660 North A’ohoku Place, Hilo, HI 96720, USA}
\altaffiltext{7}{Department of Astronomy, School of Physics, Peking University, 100871 Beijing, PR China}

\begin{abstract}
We present a pilot \hi survey of 17 Planck Galactic Cold Clumps (PGCCs) with the Five-hundred-meter Aperture Spherical radio Telescope (FAST). \hi Narrow Self-Absorption (HINSA) is an effective method to detect cold \hi being mixed with molecular hydrogen H$_2$ and improves our understanding of the atomic to molecular transition in the interstellar medium. HINSA was found in 58\% PGCCs that we observed. The  column density of HINSA was found to have an intermediate  correlation with that of $^{13}$CO, following $\rm log( N(HINSA)) = (0.52\pm 0.26) log(N_{^{13}CO}) + (10 \pm 4.1) $. \hi abundance relative to total hydrogen [\hi]/[H] has an average value of $4.4\times 10^{-3}$, which is about 2.8 times of the average value of previous HINSA surveys toward molecular clouds. For clouds with total column density N$\rm_H >5 \times 10^{20}$ cm$^{-2}$, an inverse correlation between HINSA abundance and total hydrogen column density is found, confirming the depletion of cold \hi gas during  molecular gas formation in more massive clouds. Nonthermal line width of $^{13}$CO is about 0-0.5 km s$^{-1}$ larger than that of HINSA. One possible explanation of narrower nonthermal width of HINSA is that HINSA region is smaller than that of $^{13}$CO. Based on an analytic model of H$_2$ formation and  H$_2$ dissociation by cosmic ray, we found the cloud ages to be within 10$^{6.7}$-10$^{7.0}$ yr for five sources. 

\end{abstract}
\keywords{ISM: clouds --- ISM: evolution --- ISM: molecules.}

\section{Introduction}
\label{sec:introduction}

Cold molecular cores/clumps is the cradle of star formation. Statistical properties of  prestellar cores/clumps are beneficial for understanding general evolution progress of star formation. The Planck Catalogue of Galactic cold clumps (PGCCs) released by Planck satellite \citep{2016A&A...594A..28P} contains 13188 sources with dust temperature ranging from 5.8 to 20 K. These cold clumps, identified based on the 5' beam of Planck, contain  low-mass cores to large molecular clouds, providing us a chance of investigating star formation in various scales, from 0.1 to tens of  pc.  

Spectral observations toward PGCCs provide extra kinematic and chemical properties compared to multi-band continuum observations. Molecular line surveys have been taken in CO (1-0) \citep{2012ApJ...756...76W,2012ApJS..202....4L,2016ApJS..222....7L,2018ApJS..234...28L}, C$_2$H (N=1-0) and N$_2$H$^+$ (J=1-0)  \citep{2019A&A...622A..32L}, HCO$^+$ and HCN (J=1-0) \citep{2016ApJ...820...37Y}. These molecular transitions are  collisionally excited with critical density $n_{crit}\gtrsim$ 10$^3$ cm$^{-3}$, making them tracers of dense gas. With multiple tracers, the age of a cloud can be modeled through chemical reaction networks to reproduce the observed line ratio, e.g., [C$_2$H]/[N$_2$H$^+$]  \citep{2017ApJ...836..194P,2019A&A...622A..32L}. Such estimate, however, is hardly quantitative calculation and model dependent, due to the complexities of reaction networks and incomplete knowledge of key parameters, such as densities.

\hi could be a tracer of cloud age too. When a cloud cools and evolves, the abundance of \hi in the cloud decreases with time. Cosmic ray dissociation  maintains a small content of \hi even in the central, dense region of molecular clouds, where \hi is shielded from being dissociated by UV photons and cools through collision with H$_2$ \citep{1971ApJ...165...41S, 2005ApJ...622..938G}. The cold \hi appears as absorption against its bright \hi background and correlates with molecular emission. It is called  \hi Narrow Self-Absorption (HINSA), and has been widely detected toward dark clouds in the Taurus/Perseus region and nearby molecular clouds  \citep{2003ApJ...585..823L} and molecular clouds \citep{2010ApJ...724.1402K}. HINSA is a special case of \hi  Self-Absorption (HISA), which is common and has been introduced as an effective tool to study Galactic \hi distribution  \citep[e.g.,][]{2000ApJ...540..851G, 2003ApJ...585..801D, 2003ApJ...598.1048K}. HISA  usually arises in cold neutral medium (CNM) from different mechanisms, e.g., temperature fluctuations, thus may not match with molecular emission. For the case of HINSA, it is closely associated with molecular clouds in terms of sky position, velocity, and non-thermal line width.  With HINSA analysis, a ``ring" of enhanced \hi abundance was clearly seen toward dark cloud B227 \citep{2018ApJ...867...13Z}.  

With an instantaneous effective area of 300 m diameter, the Five-hundred-meter Aperture Spherical radio Telescope \citep[FAST,][]{2011IJMPD..20..989N} located in Southern-west of China is the most sensitive radio telescope in \hi band. FAST began its commissioning after Sep 25, 2016. Its commissioning progress was reviewed in \citet{2019SCPMA..6259502J}. The 19 beam receiver was installed on FAST in mid May of 2018. Spectral backend for the 19 beam receiver was installed in late July, 2018.  

In this paper, we introduced  pilot spectral observations with 19 beam receiver of FAST. The paper is structured as follows: Observation information is described in Section \ref{sec:observation}. Analysis of deriving physical parameters is shown in Section \ref{sec:analysis}. The results and discussion are presented in Section \ref{sec:results_discussion}. We summarize our conclusion in Section \ref{sec:conclusion}.  
 
 \section{Observations and Data Archive}
 \label{sec:observation}

 \subsection{Observations of \hi}
 \label{subsec:HI}
\begin{figure*}
\centering
  \includegraphics[width=0.8\textwidth]{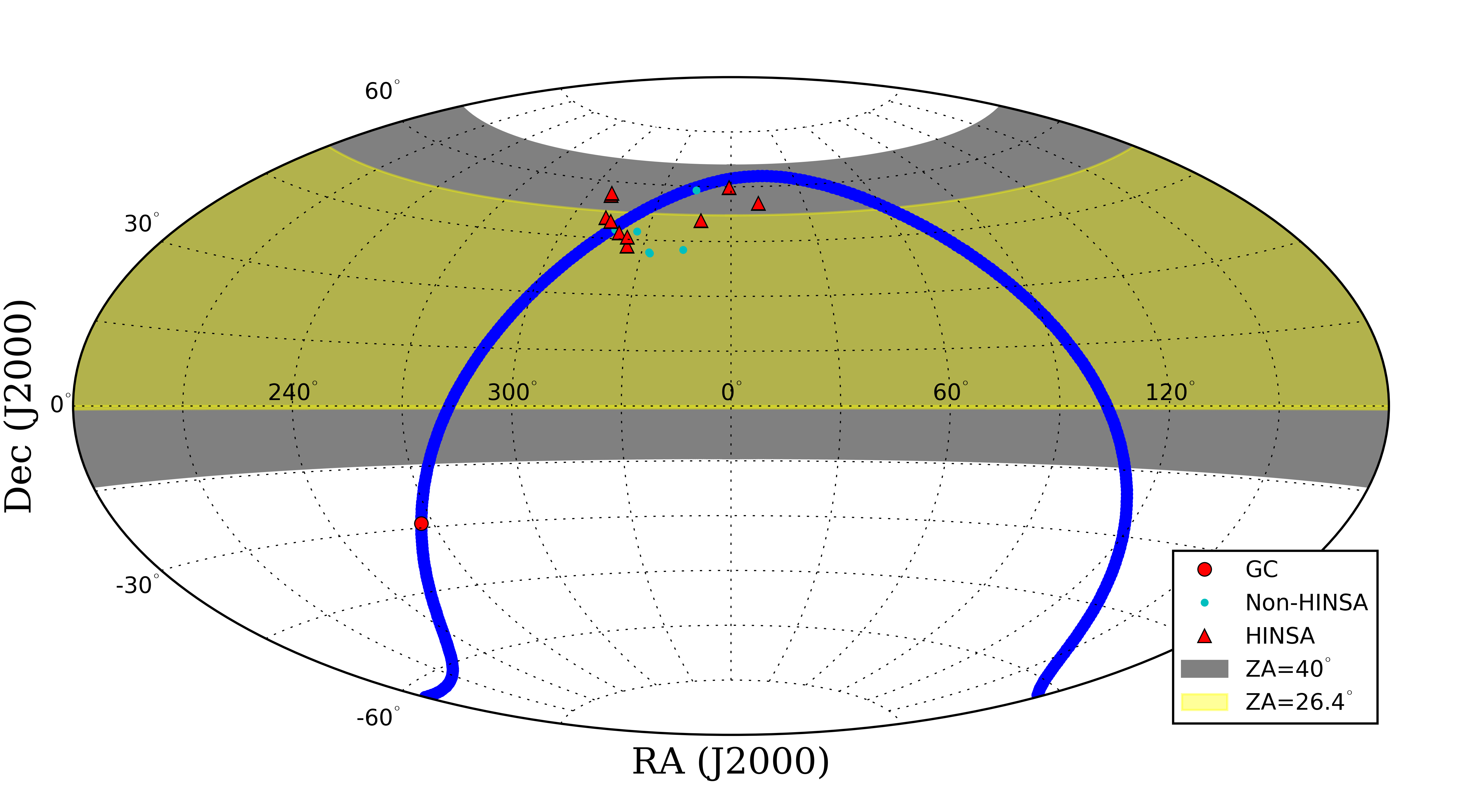}
  \caption{Spatial distribution of observed 17 PGCCs. Sources with and without HINSA detection are labeled with cyan circles and red triangles. Sky coverage of FAST within zenith angle (ZA) of 26.4 and 40 degree are shown with yellow and grey color. The Galactic plane and Galactic center (GC) are shown with a blue solid line and a red circle, respectively. } 
\label{fig:pgcc_distribution}
\end{figure*}

 We selected our sources from the catalogue of \citet{2012ApJ...756...76W} and its followup survey projects, in which more than 674 PGCCs were observed in J=1-0 transition of CO and its isotopes. Two criteria were applied: (1) position of the source is within FAST sky coverage but is out of sky coverage of Arecibo 305m telescope; (2) strong CO emission is detected toward the source. Position distribution of observed 17 PGCCs is shown in Fig. \ref{fig:pgcc_distribution}.

The FAST observations were conducted between 2018 December 16 and 2019 January 2. We observed  the  \hi\  1420.4058 MHz transition with the central beam of the 19-beam receiver on  FAST, which corresponds to a beam width (Full Width at Half Maximum; FWHM) of 2.9 arcmin.  The pointing error is $\sim$ 0.15 arcmin at 1420 MHz \citep{2019SCPMA..6259502J}. 

Single pointing  observations of PGCCs were taken with the narrow mode of spectral backend. This mode has 65536 channels in 31.25 MHz bandwidth, resulting in velocity resolution of 0.1 km s$^{-1}$ at 1.420 GHz. Each source was integrated for 10 minutes with a sampling rate of 1.0066 s. For intensity calibration, noise signal with amplitude of $\sim$ 11 K was injected under a period of 2.0132 s, which is synchronized with sampling rate. 
  
After intensity calibration, spectral  baseline was removed by applying a linear fitting of line-free range within [-300,300] \kms. During observations, system temperature was around 18 K. Reduced root mean square (rms) of the final spectrum is $\sim$35 mK, which is consistent with theoretical sensitivity calculation. FAST's current setup does not include a local oscillator (LO), and therefore neither doppler tracking nor frequency switching are available.

\begin{figure}
\centering
  \includegraphics[width=0.48\textwidth]{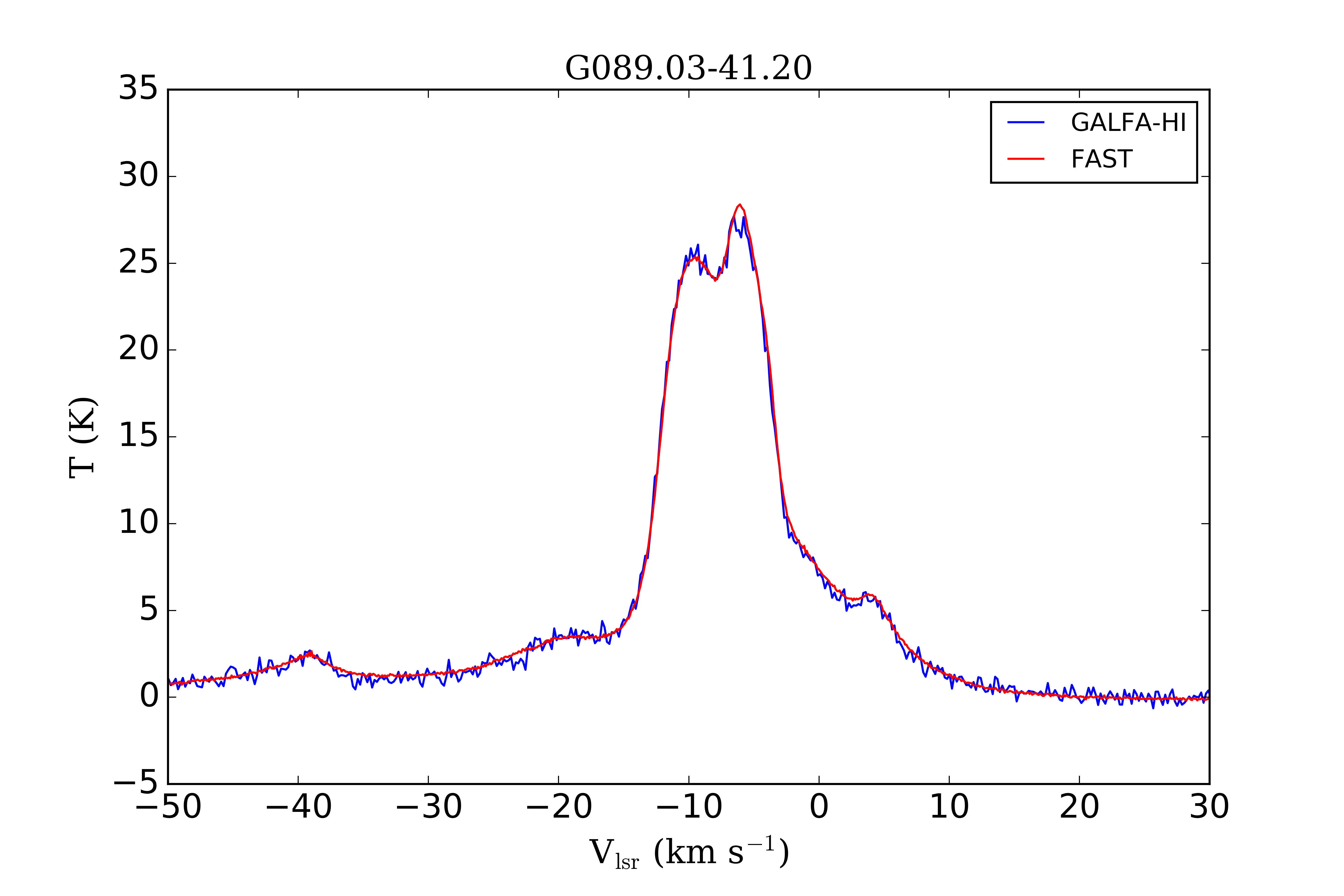}
  \caption{ \hi spectra for PGCC source G089.03-41.2.  Blue and red color represents data from GALFA-HI and FAST, respectively.  The amplitude of FAST data has been multiplied by a factor of  1.15. } 
\label{fig:hi_galfa_fast}
\end{figure}

In order to test the validity of our data reduction, the PGCC source G089.29+03.99 located within the sky coverage of Arecibo telescope was observed.  The observed spectrum with FAST is expressed in antenna temperature ($T\rm_A$) while the data of GALFA-HI  \citep{2011ApJS..194...20P, 2018ApJS..234....2P} is expressed in brightness temperature ($T\rm_{B}$).  We define a ratio $R$ to convert measured $T\rm_A$ to $T\rm_{B}$, $R=$$T\rm_{B}/$$T\rm_A$.  A value of $R=1.15$ is used for comparing observation results of G089.29+03.99. As  shown in Fig. \ref{fig:hi_galfa_fast}, we have consistent spectrum for FAST and Arecibo.   

At \hi band, telescope efficiency of FAST is almost a constant value of  $\sim$0.6 when zenith angle (ZA) is less than 26.4$^{\circ}$ and decreases linearly when ZA is greater than 26.4$^{\circ}$ \citep{2019SCPMA..6259502J}. This implies a $R$ value of 1.67 within $\rm ZA<26.4^{\circ}$.  This value is larger than  required  conversion value of 1.15 between FAST and Arecibo toward G089.29+03.99.  Three main effects may contribute to this difference.  

\begin{enumerate}
\item  The GALFA-HI map was not sampled at  the center position precisely of the PGCC source G098.29+03.99, while we pointed directly at the source.

\item The flux for the GALFA-HI map was bootstrapped using the Leiden/Argentine/Bonn (LAB) survey \citep{2005A&A...440..775K}. This means that while on a large scale the flux of the map is very accurate, there may be small-scale fluctuations.

\item   The GALFA map was partially corrected for stray radiation, while we did not.

\end{enumerate}

We applied  $R=1.67$ and corrected its dependence of ZA to  derive $T\rm_{B}$ from $T\rm_A$.

 \subsection{CO and Dust Archive Data}
 \label{subsec:CO}

The $^{12}$CO(1-0)  and $^{13}$CO(1-0)  data were obtained under single pointing mode  for investigating  dynamic evolution of PGCCs with PMO 13.7m and TRAO-14m telescopes \citep{2012ApJ...756...76W, 2016ApJS..222....7L}. Angular resolution of both PMO 13.7m and TRAO-14m is $\sim$1 arcmin (FWHM) at 115 GHz.  The backend of PMO 13.7m  provides  bandwidth of 1 GHz and  frequency resolution of 61 kHz (0.16 km s$^{-1}$ for $^{12}$CO and 0.17 km s$^{-1}$ for $^{13}$CO).  While the backend of TRAO-14m provides  bandwidth of 125 MHz and  frequency resolution of 15 kHz. This frequency resolution leads to velocity resolution of 0.04 km s$^{-1}$ at 115.271 GHz. The spectral resolution of TRAO-14m were smoothed to that of PMO 13.7m. 

The dust temperature and dust reddening toward each source were derived from Planck Collaboration et al. (2016), in which the data were smoothed into FWHM of 5 arcmin. 

\section{Analysis}
\label{sec:analysis}

We start analysis by identifying molecular components in velocity  based on multiple gaussian fitting of molecular spectrum. There are two factors:  $^{12}$CO profile may be affected by self-absorption due to its large optical depth, making it difficult to distinguish  component numbers; C$^{18}$O has a relatively low abundance compared to $^{12}$CO. Its emission is weak and may be `missed' under our observational sensitivity. Considering these factors, we identify velocity components with $^{13}$CO spectral data, whose emission is relatively optically thinner than $^{12}$CO and strong enough to be detected. A total of 23 components are identified toward 17 PGCCs.   Each component is fitted with a Gaussian function. 

Next, we check if there exists a corresponding \hi absorption feature toward each $^{13}$CO emission component.  \hi fluctuation  is common in \hi spectra. It may arise from either  kinematic effect  of two \hi clouds in different velocity or absorption of cold \hi cloud against brighter \hi background. A feature of \hi fluctuation is identified as  HINSA only when there is corresponding $^{12}$CO or $^{13}$CO emission. This method of identifying HINSA through single pointing spectrum could not absolutely exclude the coincidence between \hi fluctuation from kinematic effect and  $^{13}$CO emission.  As discussed in \citet{2000ApJ...540..851G},  geometry inspection of \hi mapping would improve this situation.  

For these identified HINSA components,  we calculated excitation temperature and column density of atomic and molecular content. They are shown as follows.

\subsection{CO Excitation}
\label{subsec:cocal}

For all observed PGCCs components,  temperature ratio  between $^{12}$CO and $^{13}$CO emission is around 2. This value is much smaller than  Galactic [$^{12}$C/$^{13}$C] ratio \citep[$\sim$ 60;][]{1998A&A...337..246L} in ISM, leading to a reasonable assumption that $^{12}$CO is optically thick, $\tau_{12} \ge 1$.  Assuming a filling factor of 1, the excitation temperature of $^{12}$CO equals  brightness temperature of $^{12}$CO. 

The excitation temperature of $^{13}$CO, $T\rm_{ex}^{^{13}CO}$ is assumed to be same of $T\rm_{ex}^{^{12}CO}$.  The optical depth of $^{13}$CO, $\tau_{13}$ is given by 
\begin{equation}
\tau_{13}= -\rm ln(1-\frac{\textit{T}_b^{13}}{\textit{T}_b^{12}}) \ ,
\label{eq:tau_13co}
\end{equation}
under the assumption of  $\tau_{12} \ge 1$. 

The column density of $^{13}$CO is derived through the following equation \citep{2002PhDT........10L,2012ApJ...760..147Q}, 
\begin{equation}
N_{\rm tot}^{\rm {^{13}CO}}=3.70\times 10^{14} \int \frac{T_{\rm b}^{13}}{K} \frac{d\upsilon}{\rm km\ s^{-1}}  f_{u}f_{\rm \tau_{13}}f_{\rm b}\frac{1}{f_{\rm beam}} \  \rm cm^{-2},
\label{eq:13co}
\end{equation}
in which $T_b^{13}$ is brightness temperature of $^{13}$, $f_{\tau_{13}}$ is the optical depth of $^{13}$CO, $f_{\tau_{13}}=\int \tau_{13} d\upsilon/\int(1-e^{-\tau_{13}})d\upsilon$. The uncertainty of $N_{\rm tot}^{\rm {^{13}CO}}$ was calculated by uncertainty propagation of spectral rms of $^{12}$CO and $^{13}$CO data. 

The above analysis of $^{13}$CO  are derived under the assumption of local thermal equilibrium (LTE).  Application of non-LTE analysis leads to a consistent  result with that of LTE for PGCC \citep{2016ApJS..222....7L}. 

\subsection{Physical Parameters  of HINSA}
\label{subsec:hical}

\citet{2003ApJ...585..823L} proposed a three-compnent model of HINSA, which consists of warm \hi background and  foreground gas with a central cold \hi cloud along line of sight.  In this model,  optical depth of background and foreground \hi gas are  $\tau_b$ and $\tau_f$, respectively.  The factor $p$ is defined as $p$=$\tau_f$/($\tau_f$+$\tau_b$).  Kinematic distance of each PGCC component was derived through Galactic  rotation curve \citep{1993A&A...275...67B}. We combine  kinematic distance  and Galactic \hi surface density distribution \citep{2003PASJ...55..191N} to estimate $p$ along sightline of the PGCC. This is different from the formula  adopted in \citet{2003ApJ...585..823L}, which is  reasonable for cloud with high Galactic latitude but would be invalid for cloud in the Galactic plane.  Details of this calculation could be found in section 4.2 of \citet{2016A&A...593A..42T}.  $T\rm_ b(\upsilon)$ and $T\rm_{\hi}(\upsilon)$  represent observed \hi brightness temperature spectrum  and brightness temperature spectrum when central cold  \hi cloud is absent, respectively.  Absorption temperature, $T\rm_{ab}(\upsilon)$ = $T\rm_{\hi}(\upsilon)-T\rm_b(\upsilon)$  could be described with \citep{2003ApJ...585..823L} ,
\begin{equation}
    T{\rm_{ab}}(\upsilon) = [pT_{\hi}(\upsilon)+(T_c-T_x)(1-\tau_f)](1-e^{-\tau}),
\end{equation}
where $T\rm_c$  is continuum temperature  including cosmic microwave background (CMB, 2.73 K), Galactic synchrotron radiation ($\sim$ 0.7 K), and continuous emission from other radiation mechanisms. The value of $T\rm_c$ is derived  by the equation $T\rm_c$ = 2.7 + $T_{c408}$($\nu\rm_{H{\textsc i}}$/408 MHz)$^{-2.8}$, in which $T_{c408}$ is continuum temperature around 408 MHz \citep{1982A&AS...47....1H}, and $\nu\rm_{H{\textsc i}}$ is  frequency of \hi emission. Derived value of  $T\rm_c$ toward our targets ranges from 3.8 to 5.6 K.  $T\rm_x$ is the excitation temperature of HINSA component, and depends on kinetic temperature T$\rm_k$ and effective temperature of radiation field. The gas of observed PGCCs  are in molecular phase, resulting a reasonable assumption that $T\rm_x=$ $T\rm_k$.  The value of   $p$ is around 0.9. We follow \citet{2003ApJ...585..823L}, the value of $\tau_f$ is assumed as 0.1, which implies a total optical depth of unity along Galactic sightline. This assumption would not significantly affect the result. The optical depth of HINSA, $\tau$ is an unknown parameter to be solved. Assuming a Gaussian broadening, $\tau$  is described in the formula of $\tau= \tau_0 e^{-\frac{(\upsilon-\upsilon_0)^2}{2\sigma^2}}$, in which $\tau_0$ is peak optical depth at central velocity.  

$T\rm_{HI}$ is recovered \hi spectrum without absorption from the central cold cloud. There are many methods to derive $T\rm_{HI}$, for instance, averaging nearby spectra without this absorption cloud \citep[e.g.,][]{2000ApJ...540..851G} or  fitting observed profile with polynomial \citep[e.g.,][]{2003ApJ...585..823L}. Considering complexity of \hi emission around the Galactic plane, the method of observing off positions or fitting with polynomial may introduce significant uncertainty. The method of polynomial fitting  depends on choice of polynomial index.  

HINSA and molecular hydrogen are believed to mix well inside the cloud.  Total velocity broadening  of each transition, $\sigma_{tot}= \sqrt{\sigma_{th}^2+\sigma_{nt}^2}$, in which $\sigma_{th}$ and $\sigma_{nt}$ represent thermal broadening due to collision  and non-thermal broadening  including turbulence, magnetic field and other effects.  \citet{2008ApJ...689..276K} introduced a new method to derive $T\rm_{HI}$ by utilizing physical  information (central velocity, non-thermal linewidth and kinematic temperature) derived from CO data to constrain fitting of  HINSA profile. We introduce this method for analysis in this paper. For each HINSA component, 4 parameters of HINSA are fitted with the following considerations \citep{2010ApJ...724.1402K}:

\begin{enumerate}
    \item Central velocity. A maximum deviation of 10\% linewidth from central velocities of $^{13}$CO is allowed. 
    \item Non-thermal linewidth. It is allowed within 50\% deviation from fitting $^{13}$CO. 
    \item Kinematic temperature, which is derived from $^{12}$CO and is kept as a constant during fitting.
    \item Optical depth. This is a relative free parameter without priori information. Initial value of 0.1 with limitation of [0,10] is adopted during fitting. 
\end{enumerate}

\begin{figure}
\centering
  \includegraphics[width=0.48\textwidth]{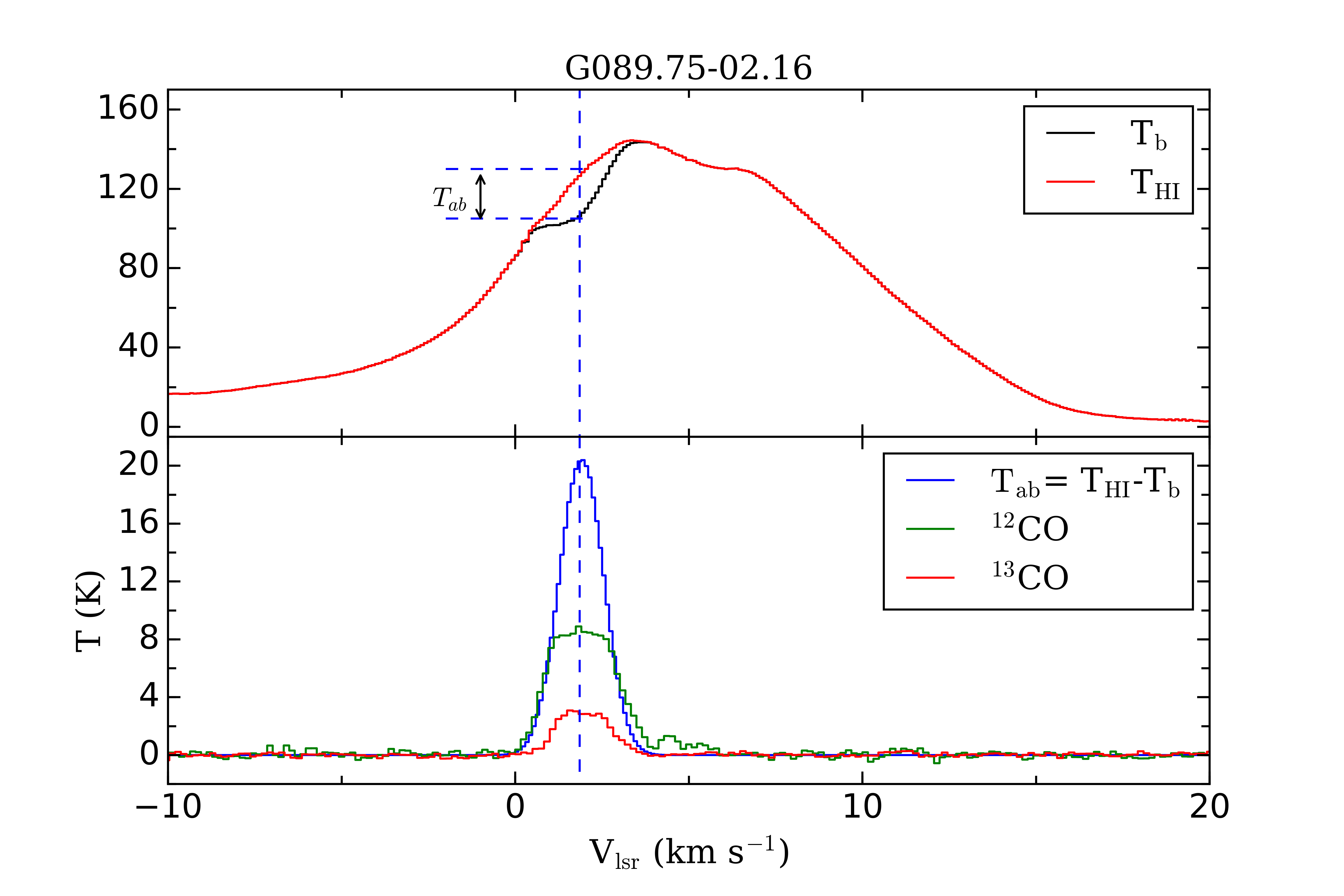}
  \caption{A example of HINSA fitting toward G089.75-02.16. $Top$ panel: black and red solid lines are the observed \hi and the recovered \hi background spectrum T$\rm_{HI}$, respectively.  $Bottom$ panel: The profile of absorption temperature $T\rm_{ab}$ (blue), $^{12}$CO (green) and $^{13}$CO spectrum (red). }
\label{fig:hinsa}
\end{figure}

\begin{figure}
\centering
  \includegraphics[width=0.48\textwidth]{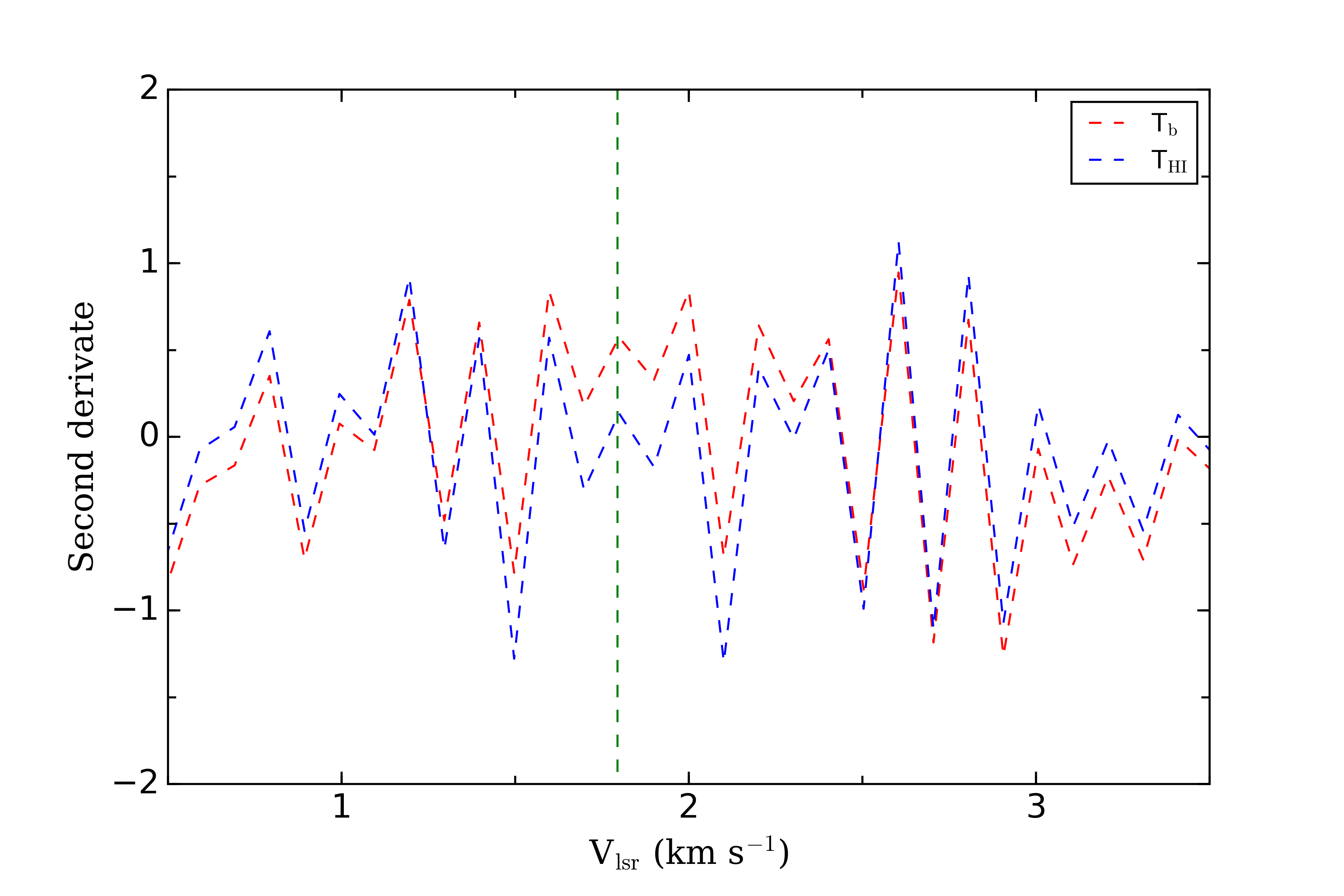}
  \caption{ Second derivatives of $T\rm_{HI}$ and $T\rm_{b}$ profile for G089.75-02.16.  The velocity range is [0.5,3.5] km s$^{-1}$.  Dashed line represents central velocity of the HINSA component. }
\label{fig:secondderive}
\end{figure}

Since the width of HINSA is narrow, HINSA feature would become apparent in  second derivative of observed spectrum \citep{2005ApJ...626..887K}. The minimization of the integrated squared sum of second derivative would be sufficient to remove HINSA feature \citep{2008ApJ...689..276K}.  We follow \citet{2008ApJ...689..276K} to derive $T\rm_{HI}$  by fitting a least minimum of second derivation of the input spectrum. This method performs perfect for simple spectrum but may lead to large deviation for complex spectrum.
The  above constraints of inputting parameters  are introduced to reduce uncertainty during fitting. An example of HINSA fitting is shown in Fig. \ref{fig:hinsa}.  Corresponding second derivatives of $T\rm_{HI}$ and $T\rm_{b}$ profile is shown in Fig. \ref{fig:secondderive}.  It is clearly seen that  second derivatives at central velocity becomes flat. Column density of HINSA is calculated by \citep{2003ApJ...585..823L}, 

\begin{equation}
N({\rm HINSA})=1.95\times10^{18}\tau_0 T_{\rm x} \Delta V\    \rm cm^{-2},
\label{eq: colhinsa}
\end{equation} 
where  $\tau_0$, $T\rm_{x}$ and $\Delta V$ are  peak optical depth, excitation temperature, and width of HINSA, respectively.  The uncertainty of N(\rm HINSA) is estimated through uncertainty propagation during fitting $\tau_0$,$T\rm_k$ and $\Delta V$.  

\subsection{Total Column Density of Hydrogen}
\label{subsec:tothydrogen}

We obtained total column  density of molecular hydrogen $N\rm_{H_2}$ and its uncertainty from \citet{2016A&A...594A..28P}.  $N\rm_{H_2}$ toward each PGCCs is derived from color extinction E(B-V), which is based on multi-band continuum emission. Obtained value of $N\rm_{H_2}$ consists contribution from all gas components along line of sight of PGCCs. For PGCCs with two or more components, we assume $N\rm_{H_2}$ of each component is linearly correlated with its column density of $^{13}$CO. The column density of hydrogen of each component is then calculated with 

\begin{equation}
    N_H = N(\rm{HINSA}) + 2N\rm_{H_2} .
\end{equation}

No estimates of $N\rm_{H_2}$ are given toward PGCCs G089.62+02.16 and G093.16+09.61.  $N\rm_{H_2}$ of gas components in these two sources are calculated from  N($^{13}$CO). Thus we have 

\begin{align}
    N_H&= N(\rm{HINSA}) + \frac{2N(\rm ^{13}CO)}{X{\rm_{^{13}CO/H_2}}} \\
        &= N(\rm{HINSA}) + \frac{2N(\rm^{13}CO)}{X{\rm_{^{13}C/^{12}C}}X{\rm_{^{12}CO/H_2}}} ,
\end{align}
where $X{\rm_{^{13}CO/H_2}}$ is $^{13}$CO/H$_2$ ratio in a cloud. It contains two parts, $X{\rm_{^{13}C/^{12}C}}$ and  $X{\rm_{^{12}CO/H_2}}$. They are $^{13}$C/$^{12}$C and $^{12}$CO/H$_2$ ratio, respectively. In previous research, the value of these two parts are generally adopted as 1/60 and 1$\times 10^4$ in Galactic ISM  \citep[e.g.,][]{2016ApJS..222....7L}. 

\begin{figure}
\centering
  \includegraphics[width=0.48\textwidth]{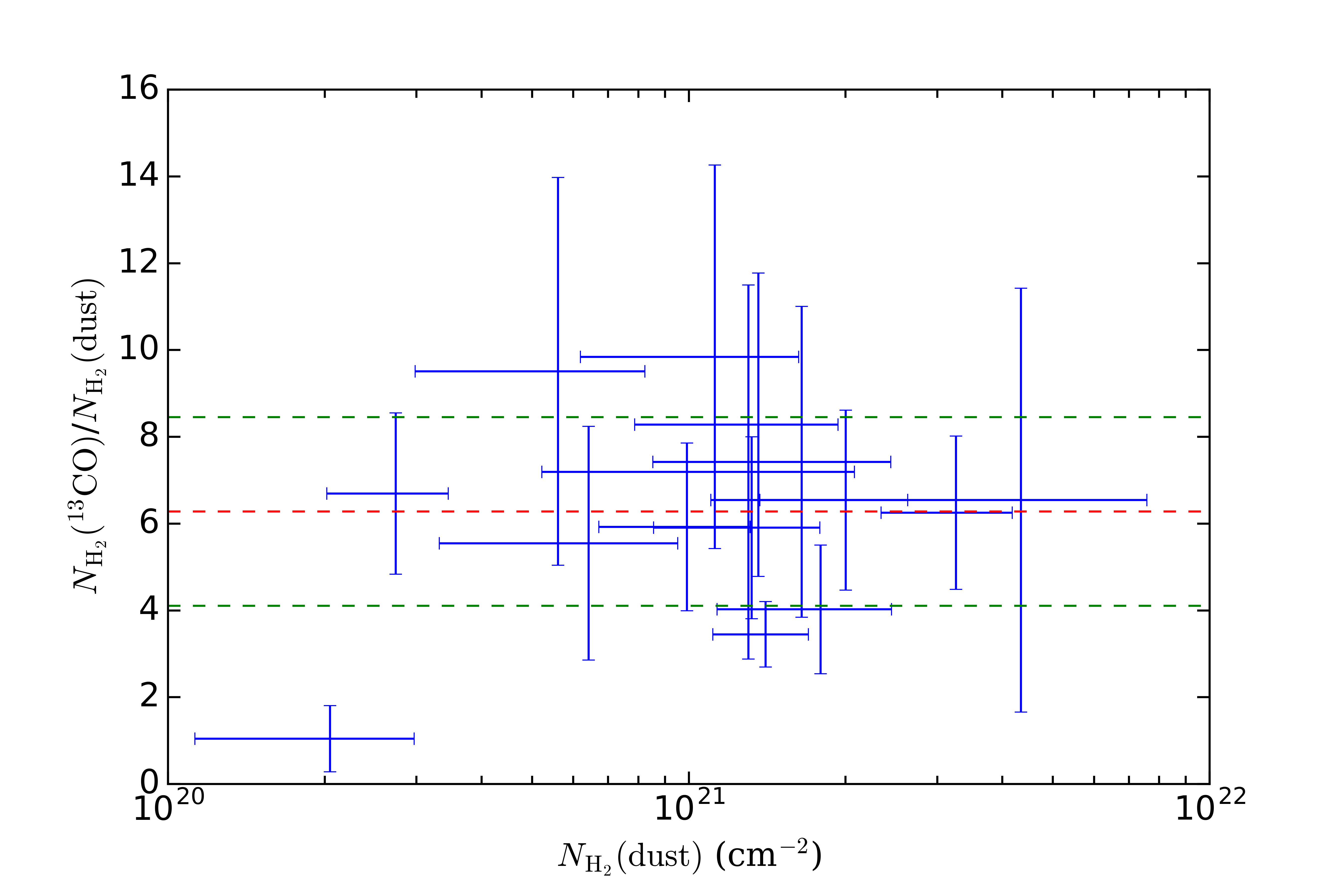}
  \caption{The $N\rm_{H_2}$($^{13}$CO)/$N_{H_2}$(dust) ratio as a function of $N\rm_{H_2}$(dust). $N\rm_{H_2}$($^{13}$CO) and $N_{H_2}$(dust) are estimated molecular hydrogen from $N(^{13}$CO) with converstion factor of 1.67$\times 10^{-6}$ and dust (Planck collaboration 2016), respectively.  Red and green dashed line represents averaged value and 1 $\sigma$ deviation, respectively.}
\label{fig:NH2_comapre}
\end{figure}

We compare validity of calculating $N(\rm H_2)$ by adopting $X{\rm_{^{13}CO/H_2}}=1.67\times 10^{-6}$ for observed PGCCs with $N(\rm H_2)$ estimation from dust. As shown in Fig. \ref{fig:NH2_comapre}, the adoption of $X{\rm_{^{13}CO/H_2}}=1.67\times 10^{-6}$ would overestimate $N\rm_{H_2}$ by a factor of 6.2. This discrepancy may origin from variation of $^{12}$C and $^{13}$C in ISM with different environment. We adopt $X{\rm_{^{13}CO/H_2}}= (1.05\pm 0.36)\times 10^{-5}$ for calculating $N\rm_H$ in PGCCs G089.62+02.16 and G093.16+09.61.

\section{Results and Discussion}
\label{sec:results_discussion}

A total of 17 PGCCs inside FAST sky coverage were surveyed. We have detected 13 HINSA components toward 10 PGCCs, implying a HINSA detection rate of $\sim$58\% in PGCCs. This detection percentage is smaller than that of 77\% in Taurus/Perseus region  \citep[hereafter LG03]{2003ApJ...585..823L} and 80\% of 48 molecular clouds \citep[hereafter KG10]{2010ApJ...724.1402K}. This deviation may arise from limited sample number and different strategy of sample selection. 

We unbiasedly selected PGCCs without constraint on visual extinction and presence of Young Stellar Objects (YSOs), which could prevent cold gas from external and internal heating. A embedded YSO is found to locate inside PGCCs G108.84-00.81. Though LG03 stated that the presence of YSOs would not significantly affect the detection rate of HINSA, the statement should be checked further by including  molecular clouds with far distance. In KG10, they selected sample clouds with peak visual extinction $A\rm_V >$ 6 mag and without YSOs inside. 

The spectra of sources with HINSA detection and without HINSA detection are shown  in Appendix \ref{sec:hinsa_all} and Appendix \ref{sec:nonhinsa_all}, respectively. There are many reasons that may be responsible for non-HINSA detection. For example, the excitation temperature of the HI gas inside a molecular cloud is comparable to that of background continuum temperature.  As shown in Fig. \ref{fig:pgcc_distribution}, there is no obvious spatial difference between HINSA  and non-HINSA detection. The sample of 17 PGGCs are small. More samples are needed to further investigate this point.  

\subsection{HINSA Column Density and Abundance}
\label{subsec:detectionrate}

N(HINSA) of PGCCs has an average of 6.3$\times$ 10$^{18}$ cm$^{-2}$. As a comparision,  average value of N(HINSA) in LG03 and KG10 are 7.2$\times$ $10^{18}$ and 6.3$\times$ 10$^{18}$ cm$^{-2}$. These results are consistent though we focus on different category of molecular cloud.

HINSA abundance, N(HINSA)/N$\rm_H$ represents abundance of cold \hi in a cloud. The average HINSA abundance in PGCCs is $4.4\times 10^{-3}$, which is about 3 times the value derived in both LG03 and KG10.  We notice that $N_{H_2}$ is derived  from Planck data with 5' FWHM, which is larger than that of  \hi (2.9') and CO (1') data we used.  This  would lead to overestimation of  N(HINSA)/N$\rm_H$.  As an estimation,  the  peak column density $N\rm_{H_2}^{peak}$ of PGCC would be 1.38 times the value of $N\rm_{H_2}$  \citep{2016A&A...594A..28P}.  Considering this uncertainty, the lower limit of averaged N(HINSA)/N$\rm_H$ would be $3.2\times 10^{-3}$. 

\begin{figure}
\centering
  \includegraphics[width=0.48\textwidth]{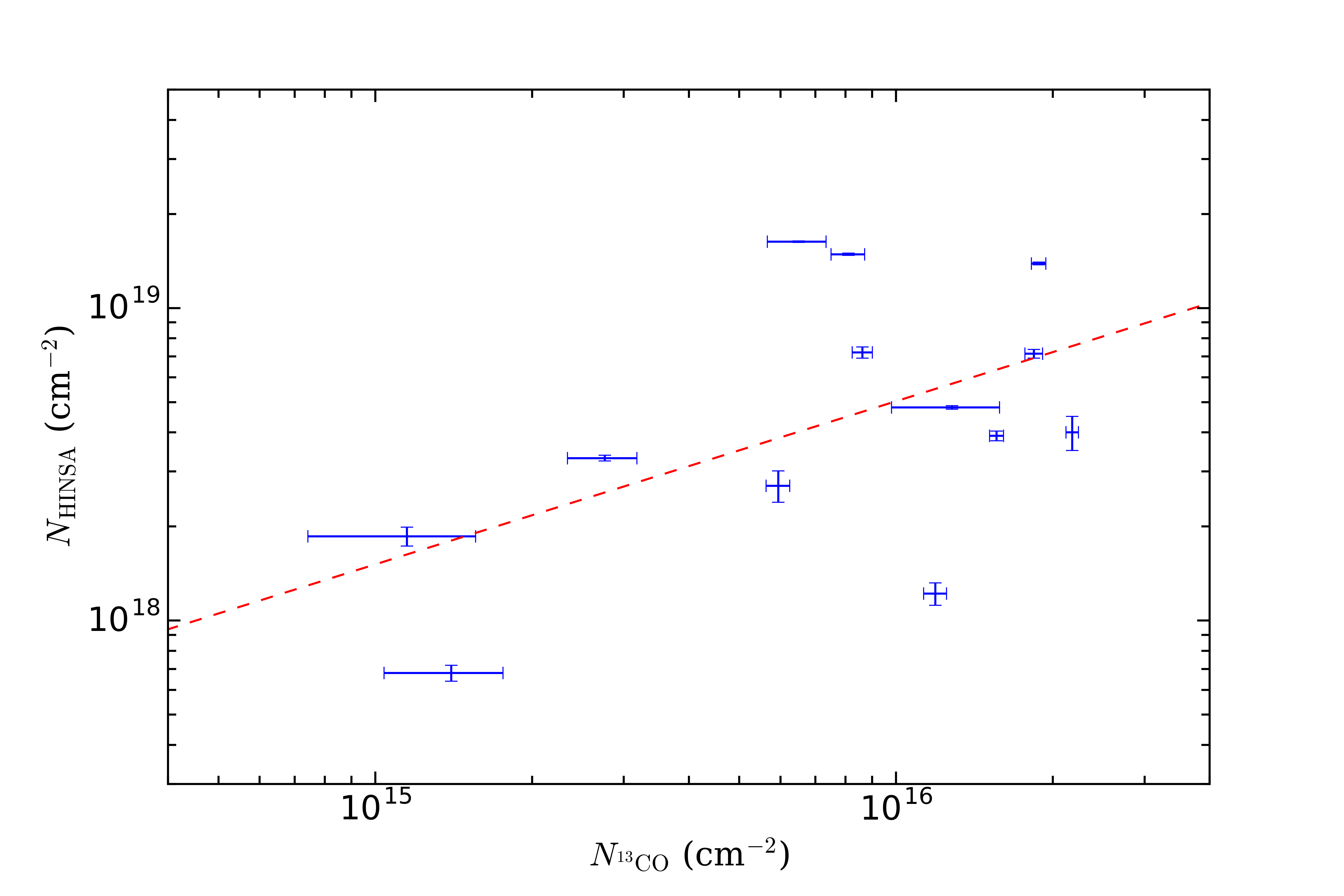}
  \caption{Correlation between column density of HINSA and $^{13}$CO.  The red dashed line represents linear fitting of the data. }
\label{fig:hinsa_13co}
\end{figure}

The correlation between log($N(\rm HINSA)$) and log($N(\rm ^{13}CO)$) is investigated in Fig. \ref{fig:hinsa_13co}.  An extra  uncertainty of 20\% is considered  due to flux calibration of $^{13}$CO. As described in section \ref{subsec:hical}, the uncertainty of fitting complex spectrum is difficult to estimate quantitatively and  is not included in the error bar of N(HINSA) of Fig. \ref{fig:hinsa_13co}. The correlation can be expressed with 
\begin{equation}
\rm log( N(HINSA)) = a\  log(N_{^{13}CO}) +b,
\label{eq:hinsa_13co_fitting}
\end{equation}
in which a= 0.52$\pm$0.26 and b=10$\pm$4.1. The Pearson correlation coefficient is 0.52, indicating an intermediate  correlation between HINSA and $^{13}$CO. 

The uncertainty of $N\rm_{H_2}$ from dust could reach 50\% the value itself. It is larger than the uncertainty percent of N(HINSA) and $N(\rm ^{13}CO)$.  The uncertainties of  HINSA abundance N(HINSA)/N$\rm_H$ and total column density N$\rm_H$ are significant and  must be included during correlation analysis.  When uncertainty in both X and Y axes  are considered, the correlation between N(HINSA)/N$\rm_H$ and N$\rm_H$ are fitted with the following equation, 
\begin{equation}
\rm log( N(HINSA)/N_H) = c\  log(N_H) +d.
\label{eq:hinsa_NH_fitting}
\end{equation}

\begin{figure}
\centering
  \includegraphics[width=0.48\textwidth]{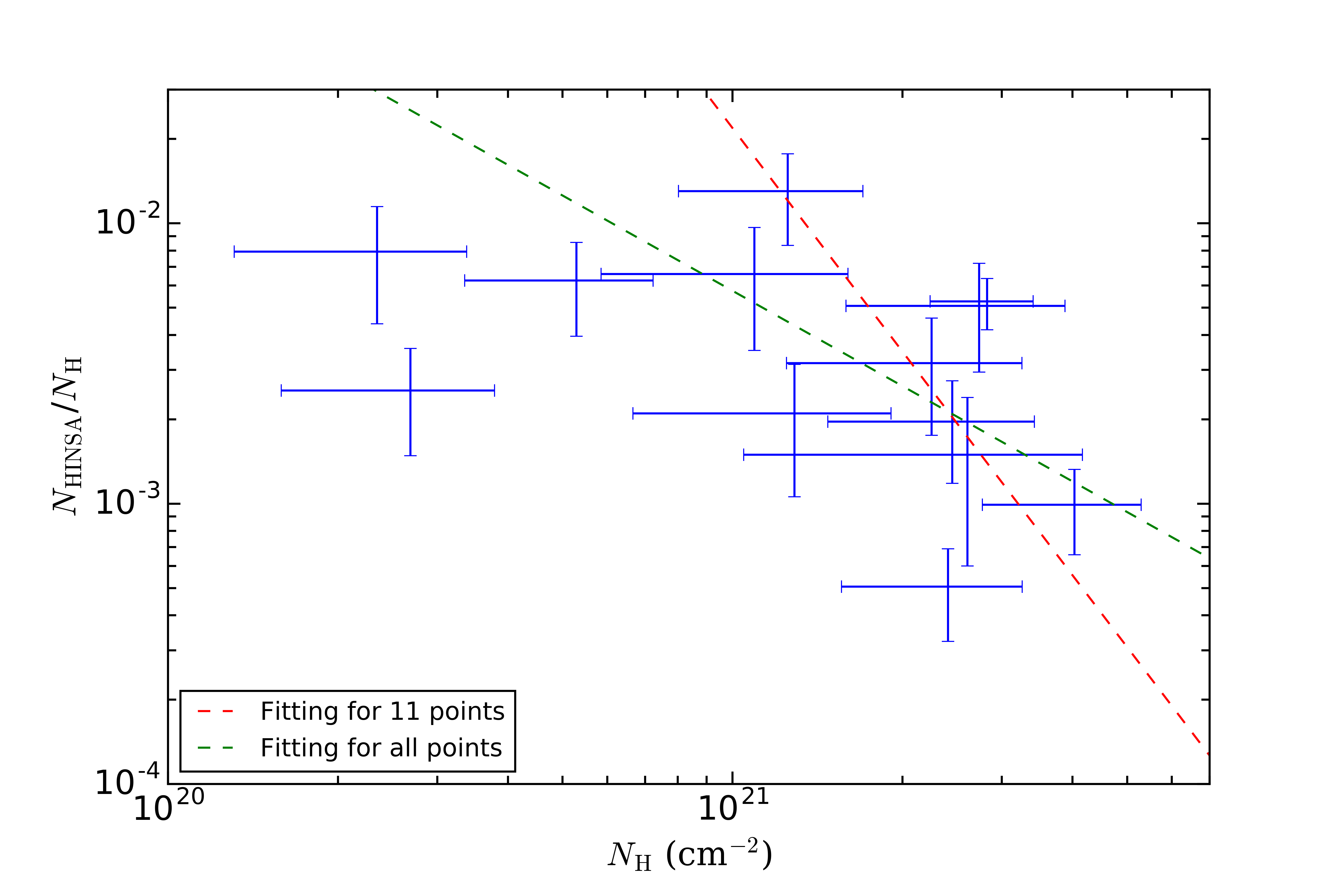}
  \caption{The relation between HINSA abundance $N\rm_{HINSA}$/$N\rm_H$ compared to total hydrogen $N\rm_H$.  The red and green dashed lines represent fitting for the right 11 data points and all 13 data points. }
\label{fig:hinsa_abundance}
\end{figure}

As shown in  Fig. \ref{fig:hinsa_abundance}, there is a trend that  HINSA abundance decreases along increasing hydrogen column density. The fitted parameter c =  -1.1$\pm$ 0.32 and  d=21$\pm$ 6.7 when all data points are included. The slope of  -1.1$\pm$ 0.32 could not exclude the possibility that linear decrease of  HINSA abundance is just due to division of N$\rm_H$. The two components, -3.73  km s$^{-1}$ of G093.16+09.61 and  4.54 km s$^{-1}$ of  G107.64-09.31 show significant deviation.  The $^{13}$CO intensity of these two components are low, which may lead to underestimation of $N$($^{13}$CO), and thus underestimation of N$\rm_H$ with the calculation method we used in this paper. When these two components are excluded, the fitted parameter c =  -2.7$\pm$ 1.0 and  d= 54$\pm$ 22. The slope of  -2.7$\pm$ 1.0 indicates  that the decreasing trend is physically meaningful. HINSA column density will decrease in increasing N$\rm_H$. This result implies that cold \hi would be depleted in denser molecular cloud.  Since the two components with low $^{13}$CO emission would affect this analysis, more HINSA samples with accurate estimation of N$\rm_H$ are needed to check this correlation. 

\subsection{Kinematic Environment}
\label{subsec:detectionrate}

\begin{figure}
\centering
  \includegraphics[width=0.48\textwidth]{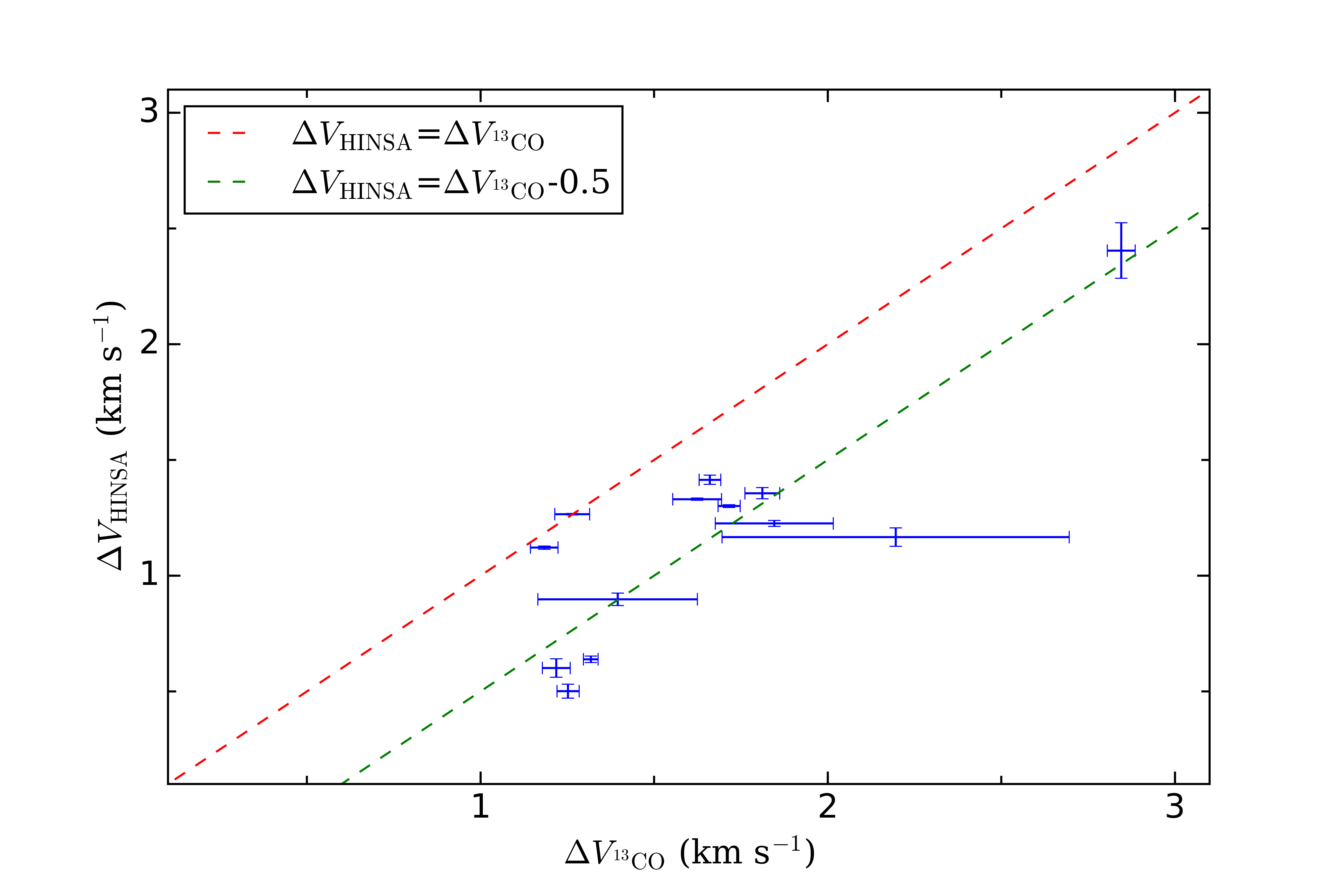}
  \caption{Nonthermal width of HINSA and $^{13}$CO. }
\label{fig:deltav}
\end{figure}

We investigated further about kinematic environment of PGCCs. Thermal and nonthermal effect (e.g., turbulence) are two main factors that lead to line broadening. Since PGCCs are cold ($\sim$ 5-25 K), the thermal broadening width is small, $\sim$ 0.3 km s$^{-1}$. Unlike thermal effect, nonthermal effect does not depend on kinetic temperature and particle mass. In order to check the turbulence effect on different species, We compare nonthermal width of HINSA and CO  in Fig. \ref{fig:deltav}. We found that nonthermal width of $^{13}$CO is about  0-0.5 km s$^{-2}$ larger than that of HINSA  though  spatial resolution of CO ($\sim$ 1 arcmin) is smaller than that of  \hi ($\sim$ 3 arcmin).  The smoothing of CO data to same resolution of \hi is expect to result in larger linewidth of $^{13}$CO, enlarging disparity between CO and \hi linewidth. 

In order to check if this result is due to applying constraints of non-thermal width while fitting HINSA in  Section \ref{subsec:hical}, a wider limitation was chosen. We adopted  100\%  deviation from fitting  $^{13}$CO. Variation of fitted line width under new limitation is less than 0.5\% compared to old one.  Limitation of line width is used for deriving local extreme solution and does not affect fitting result. 

Line broadening caused by turbulence ($\Delta V$) is found to connect with cloud size ($L$) through a relation , $\Delta V\sim L^q$. The value of q is 0.38 when first proposed by \citet{1981MNRAS.194..809L}. Further observations toward giant molecular clouds indicate a value of 0.5 \citep[e.g.,][]{1987ApJ...319..730S}. This relation implies that smaller regions are less turbulent.  The broader line width of $^{13}$CO than HINSA may indicate that HINSA lie in a smaller, interior region than $^{13}$CO.  

\subsection{Transition between Atomic and Molecular Phase}
\label{subsec:atomic2molecular}

The existence of cold \hi in clumps is due to balance between H$_2$ formation on dust grain and H$_2$ dissociation by external environment, e.g., illumination from cosmic ray. HINSA abundance can be an effective indicator of cloud evolution \citep{2005ApJ...622..938G}. In order to compare with analytic model, we have to identify volume density of total hydrogen. For PGCCs with both distance estimation $d$ (including kinematic distance derived rotation curve) and cloud size in angular resolution $\Delta\theta$, the cloud size is estimated with $L=d\Delta\theta$. For other PGCCs, we adopt the fitting formula derived in  \citet{2004ApJ...615L..45H}, $\sigma =0.9(L/1\rm pc)^{0.56}$ km s$^{-1}$ to calculate the size of PGCCs. Average density is calculated with $\bar{n}\rm{_H}=\frac{N\rm{_H}}{L}$. 

We follow the analytic model in \citet{2005ApJ...622..938G}. In a cloud with \hi density of $n\rm_{\hi}$ and H$_2$ density of  $n\rm_{H_2}$,  its \hi density fraction $x_1= n\rm_{\hi}/$$(n\rm_{\hi}+$$2n\rm_{H_2})$ as a function of time $t$ is given by
\begin{equation}
x_1(t) = 1- \frac{2k'n_0}{2k'n_0+\zeta_{H_2}}[1-\textrm{exp}(\frac{-t}{\tau\rm_{HI\rightarrow H_2}})],
\label{eq:hi_fraction}
\end{equation}
in which $k'$ is the formation rate coefficient of H$_2$.  $\zeta\rm_{H_2}$ is the ionization rate of H$_2$ by cosmic-ray. The value of $5.2\times 10^{-17}$ s$^{-1}$ has been adopted \citep{2000A&A...358L..79V}. Time constant between atomic and molecular phase $\tau\rm_{HI\rightarrow H_2}=$$\frac{1}{2k'n_0+\zeta\rm_{H_2}}$.  By adopting  $x_1=$$N(\rm HINSA)/N\rm_H$ and   $n_0=\bar{n}\rm{_H}$, cloud age could be calculated from Equation \ref{eq:hi_fraction}.

\begin{figure}
\centering
  \includegraphics[width=0.48\textwidth]{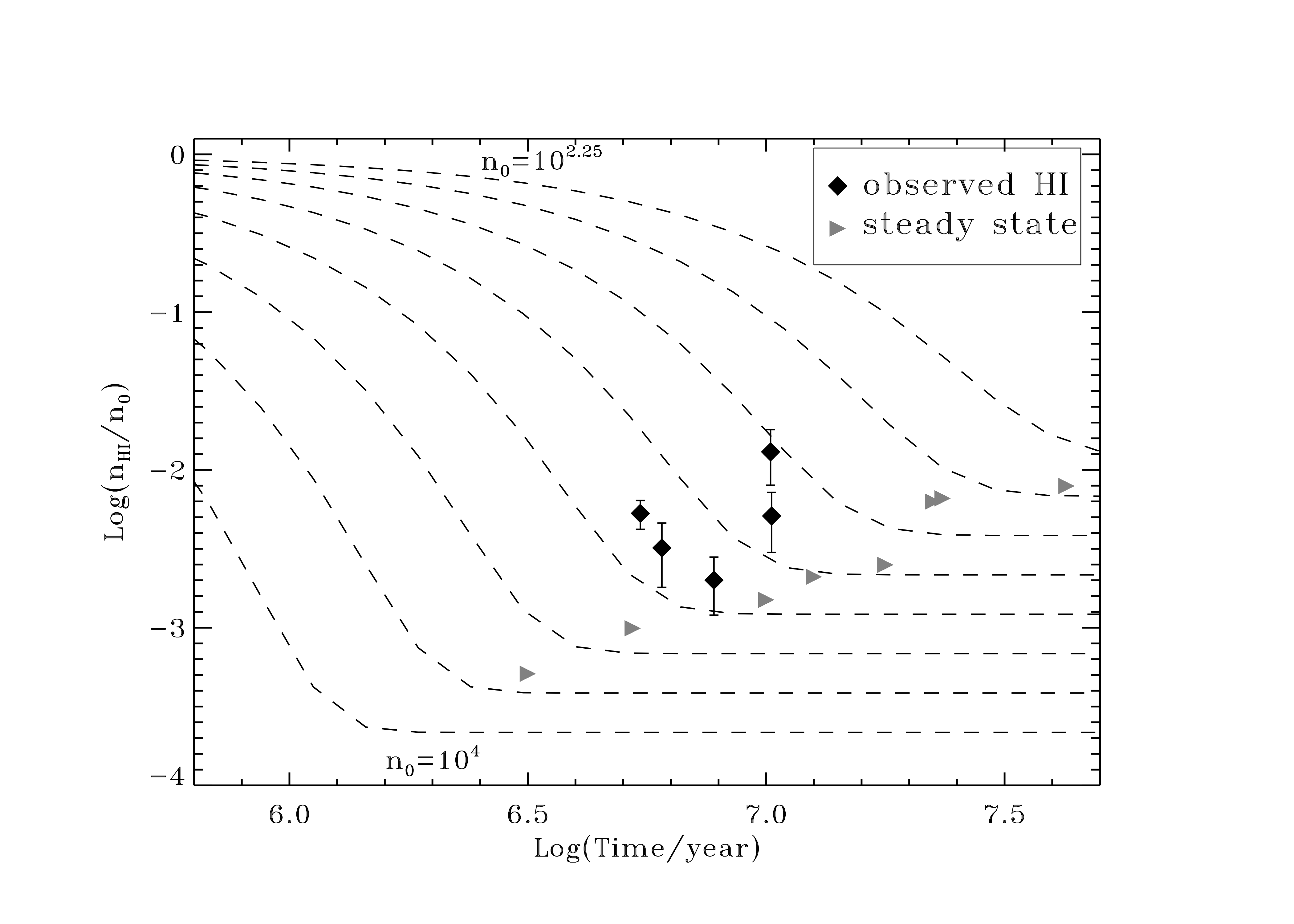}
  \caption{ Time dependence of \hi abundance for various proton density (in cm$^{-3}$). The dashed lines represent different logarithmic densities with index interval of 0.25. Five sources are plotted with black diamonds.  The cloud age of rest 8 sources is calculated  with measured \hi abundance and the assumption of steady state. They are shown with grey triangles. }
\label{fig:age_evolution}
\end{figure}

Solution can be derived for only 5 of 13 PGCC components. Fig. \ref{fig:age_evolution} shows the evolution age as a function of abundance of \hi and total proton density. Cloud age of these 5 clouds lies in the range $10^{6.7}$ yr $\leqslant$ $\tau\rm_{cloud}$ $\leqslant$ $10^{7.0}$ yr. Besides, they locate in a transition phase between unstable and steady state.  
 
Cloud has density profile distribution with cloud radius, e.g., power law  density profile \citep{1977ApJ...214..488S}, which results in a denser central region.  The beam size of  \hi  ($\sim$ 2.9 arcmin) and CO ($\sim$ 1 arcmin) observations  allows us to only cover  the central dense region of a PGCC clump.  The estimation of cloud density by simply dividing cloud size with column density would underestimate cloud density, leading to overestimate cloud age.  As discussed in section \ref{subsec:detectionrate}, we adopt peak proton density, which is 1.38 times the current value  to estimate a lower limit of cloud age. Lower limits of cloud age for 5 PGCC components is in the range $10^{6.6}$ yr $\leqslant$ $\tau\rm_{cloud}$ $\leqslant$ $10^{6.9}$ yr. These value are about 23\% smaller than that shown in Fig. \ref{fig:age_evolution}.
 
We are not able to find available solution for the rest 8 components due to relatively low \hi abundance under its density. A possible reason is that cloud size was overestimated, resulting underestimation of total hydrogen density.

\section{Summary}
\label{sec:conclusion}

In this paper, we have observed \hi spectra toward 17 PGCCs with FAST. Thirteen HINSA components are found toward 10 sources.  The abundance of cold \hi in these clouds varies from $5.1\times 10^{-4}$ to $1.3\times10^{-2}$ with an average value of $4.4\times 10^{-3}$. 

We found that HINSA column density in PGCCs has a intermediate correlation with $^{13}$CO  column density at logarithmic scale. This correlation follows a relationship of $\rm log( N(HINSA)) = (0.52\pm 0.26) log(N_{^{13}CO}) + (10 \pm 4.1) $.

When all date points are included, we could not exclude the possibility that decreasing HINSA abundance (N(HINSA)/N$\rm_H$) is due to division of N$\rm_H$.  If we only consider HINSA components with $N_H$> $5\times 10^{20}$ cm$^{-2}$,   decreasing HINSA abundance  as a function of  $N\rm_H$ is physically meaningful. This confirms depletion of cold \hi for forming molecular gas in more massive PGCCs. 

The nonthermal linewidths of 11 of 13 HINSA components are narrower  than that of $^{13}$CO. This result may implicate that HINSA  located in inner and smaller region than that of $^{13}$CO. 

We compare our results with analytic model of cloud evolution. Five of 13 HINSA components satisfy cloud evolution at age range of 10$^{6.7}$- 10$^{7.0}$ yr.

Due to a limited sample of this pilot survey with FAST, there still exist some uncertainties in the above results. A larger survey of HINSA features are necessary for improving these results.  

\normalem
\begin{acknowledgements}

We are grateful to the anonymous referee for his/her constructive suggestions, which have greatly improved this paper. This work is supported by the National Natural Science Foundation of China grant No. 11988101, No. 11725313 and No. 11803051, National Key R\&D Program of China No. 2017YFA0402600,  CAS International Partnership Program No.114A11KYSB20160008, National Science Foundation of China (11721303) and the National Key R\&D Program of China (2016YFA0400702), CAS "Light of West China" Program, CAS Interdisciplinary Innovation Team and Young Researcher Grant of National Astronomical Observatories, Chinese Academy of Sciences. LQ is supported in part by the Youth Innovation Promotion Association of CAS (id.~2018075).

\end{acknowledgements}


\begin{sidewaystable}
\fontsize{7}{6}\selectfont
\centering
\caption{Derived parameters of observed sources. \label{table:source}}
\renewcommand{\arraystretch}{1.5}
\begin{tabular}{lllccccccccccc}
\hline
\hline
\multirow{3}{*}{Source}  & \multicolumn{1}{c}{\multirow{3}{*}{RA(J2000)}}  &\multicolumn{1}{c}{\multirow{3}{*}{Dec(J2000)}}  &
\multicolumn{1}{c}{\multirow{3}{*}{Comp}}  &\multicolumn{1}{c}{$^{12}$CO} & \multicolumn{4}{c}{$^{13}$CO} & \multicolumn{4}{c}{HINSA}  &  \multicolumn{1}{c}{\multirow{3}{*}{N(H)}}   \\
 \cline{5-5}
 \cline{7-9}
 \cline{11-13}
 
 & & & & \multicolumn{1}{c}{$T\rm_{ex}$} & & \multicolumn{1}{c}{V$_{cen}$} & \multicolumn{1}{c}{$\Delta V$} & \multicolumn{1}{c}{$N\rm_{^{13}CO}$} & & \multicolumn{1}{c}{V$_{cen}$} & \multicolumn{1}{c}{$\Delta V$} & \multicolumn{1}{c}{$N\rm_{HINSA}$} &  \\
 
   &&&& \multicolumn{1}{c}{(K)}& &\multicolumn{1}{c}{(km s$^{-1}$)}& \multicolumn{1}{c}{(km s$^{-1}$)}& \multicolumn{1}{c}{(cm$^{-2}$)}&& \multicolumn{1}{c}{(km s$^{-1}$)}& \multicolumn{1}{c}{(km s$^{-1}$)}& \multicolumn{1}{c}{(cm$^{-2}$)}&  \multicolumn{1}{c}{(cm$^{-2}$)} \\
\hline
 G089.29+03.99 & 20:50:43.20   &  50:25:38.7 & 1 & 15.95$\pm$0.31  & & -4.41$\pm$0.02  &  2.85$\pm$0.04  &  2.18$\pm$0.05  & & -4.70$\pm$0.02  &  2.48$\pm$0.12  &  4.0$\pm$0.5  &  20.15$\pm$6.32\\
 G089.62+02.16 & 21:00:44.18   &  49:30:29.8 & 1&  8.47$\pm$0.87  & & -1.39$\pm$0.44  &  1.85$\pm$0.17  &  0.28$\pm$0.04  & & -1.21$\pm$0.02  &  1.30$\pm$0.01  &  3.31$\pm$0.07  &  2.63$\pm$0.97 \\  
 &     &    & 2 &  9.08$\pm$0.86  & & -0.16$\pm$0.07  &  1.27$\pm$0.05  &  0.65$\pm$0.07  & & -0.29$\pm$0.02  &  1.35$\pm$0.003  &  16.31$\pm$0.09  &  6.18$\pm$2.25  \\
 G089.64-06.84 & 21:37:58.08   &  43:11:7.8 & 1 &   18.07$\pm$0.27  & & 12.86$\pm$0.01  &  1.33$\pm$0.02  &  1.56$\pm$0.04  & & 12.77$\pm$0.01  &  0.91$\pm$0.01  &  3.90$\pm$0.14  &  13.01$\pm$7.79  \\
 G089.75-02.16 & 21:20:11.99   &  46:38:41.3 & 1 &   20.99$\pm$0.39  & & 1.86$\pm$0.02  &  1.73$\pm$0.03  &  1.88$\pm$0.05  & & 1.79$\pm$0.01  &  1.48$\pm$0.01  &  13.88$\pm$0.14  &  13.61$\pm$5.73  \\
 G090.76-04.57 &   21:34:00.72  &  45:36:504 & 1 &   17.81$\pm$0.24  & & 10.13$\pm$0.02  &  1.82$\pm$0.05  &  0.86$\pm$0.03  & & 10.21$\pm$0.01  &  1.50$\pm$0.02  &  7.21$\pm$0.30  &  5.43$\pm$2.54  \\
 G092.79+09.12 & 20:37:48.71   &  56:19:27.1 & 1 &  11.8$\pm$0.5  & & -2.53$\pm$0.03  &  1.63$\pm$0.07  &  0.81$\pm$0.05  & & -2.37$\pm$0.01  &  1.43$\pm$0.01  &  14.86$\pm$0.13  &  14.04$\pm$2.93 \\ 
 G093.16+09.61 & 20:36:31.44   &  56:54:48.6 & 1&  8.43$\pm$0.49  & & -3.73$\pm$0.10  &  1.40$\pm$0.23  &  0.140$\pm$0.030  & & -3.61$\pm$0.01  &  1.00$\pm$0.03  &  0.68$\pm$0.04  &  1.34$\pm$0.55 \\
  &    &  & 2 &   9.19$\pm$0.49  & & -1.75$\pm$0.02  &  1.19$\pm$0.04  &  1.28$\pm$0.25  & & -1.68$\pm$0.01  &  1.21$\pm$0.01  &  4.81$\pm$0.06  &  12.23$\pm$4.88  \\
 G107.64-09.31 & 23:14:26.40   &  50:39:44.6 & 1&   10.9$\pm$0.4  & & 4.54$\pm$0.22  &  2.20$\pm$0.50  &  0.115$\pm$0.034  & & 4.72$\pm$0.02  &  1.27$\pm$0.04  &  1.86$\pm$0.13  &  1.16$\pm$0.51  \\
  &    &   & 2&   16.3$\pm$0.4  & & 6.98$\pm$0.02  &  1.26$\pm$0.03  &  1.19$\pm$0.05  & & 6.85$\pm$0.02  &  0.79$\pm$0.03  &  1.22$\pm$0.10  &  12.04$\pm$4.25  \\
 G116.08-02.40 & 23:56:45.11   &  59:44:20.0 & 1&   16.52$\pm$0.32  & & -0.96$\pm$0.01  &  1.67$\pm$0.03  &  1.84$\pm$0.06  & & -1.00$\pm$0.01  &  1.54$\pm$0.02  &  7.14$\pm$0.23  &  11.22$\pm$5.02  \\
 G122.01-07.47 & 00:45:00.02 &  55:23:18.6 & 1 &  17.17$\pm$0.22  & & -52.18$\pm$0.02  &  1.23$\pm$0.04  &  0.59$\pm$0.03  & & -52.21$\pm$0.02  &  0.87$\pm$0.04  &  2.7$\pm$0.3  &  6.42$\pm$3.10  \\
 \hline
 G089.36-00.69 & 21:12:24.00   &  47:23:32.3 & 1&   7.43$\pm$0.28  & &  -24.01$\pm$0.07 &  1.85$\pm$0.17  &  0.155$\pm$0.020 &&-&-&-& 2.97$\pm$ 0.84 \\
  &    &   & 2&   17.77$\pm$0.27 &  &    2.29$\pm$0.02 &  2.71$\pm$0.03  &   3.19$\pm$0.05 &&-&-&-& 61.23$\pm$ 17.26 \\
  &    &  & 3 &   8.89$\pm$0.28  & &    4.94$\pm$0.41 &  1.93$\pm$0.56  &  0.052$\pm$0.020 &&-&-&-& 1.00$\pm$ 0.28 \\
 G090.76-04.57 &   21:34:00.72  &  45:36:504 & 2 &   7.31$\pm$0.25  & & 12.64$\pm$0.06  &  0.47$\pm$0.14  &  0.028$\pm$0.011  &&-&-&-&  0.18$\pm$0.11  \\
 G093.22-04.60 & 21:44:44.88   &  47:13:3.4 & 1&    17.16$\pm$0.25  & &    3.60$\pm$0.01 &  1.52$\pm$0.02  &   2.04$\pm$0.04 &&-&-&-& 32.9$\pm$ 15.8 \\
 G093.31-11.68 &  22:09:44.88  &  41:42:30.2 & 1&    15.32$\pm$0.39 & &   -0.35$\pm$0.02 &  0.83$\pm$0.05  &  0.305$\pm$0.028 &&-&-&-& 5.47$\pm$1.43 \\
 G093.36-12.07 &  22:11:11.52  &  41:25:35.8 & 1&    8.52$\pm$0.37 & &    0.24$\pm$0.16 &  0.86$\pm$0.37  &  0.036$\pm$0.020 &&-&-&-& 4.09$\pm$ 1.84 \\
 G101.14-15.29 &  22:54:59.75  &  42:35:51.7 & 1&    15.45$\pm$0.43 & &   -9.59$\pm$0.01 &  0.97$\pm$0.03  &   0.98$\pm$0.05 &&-&-&-& 18.82$\pm$ 6.39 \\
 G108.84-00.81 &  22:58:48.72  & 58:56:54.6  & 1&  22.74$\pm$0.45 & &  -49.67$\pm$0.02 &  3.27$\pm$0.04  &   4.73$\pm$0.10 &&-&-&-& 86.8$\pm$ 64.3 \\
 G121.93-07.67 &  00:44:28.34 & 55:11:24.4  & 1&   17.01$\pm$0.21 & &  -52.07$\pm$0.01 &  1.18$\pm$0.01  &  1.202$\pm$0.024 &&-&-&-& 35.8$\pm$ 13.2 \\

\hline
\end{tabular}
\end{sidewaystable}


\clearpage
\appendix

\section{Detected HINSA toward PGCCs}
\label{sec:hinsa_all}
\begin{figure}[!htp]
\centering
\includegraphics[width=0.4\textwidth]{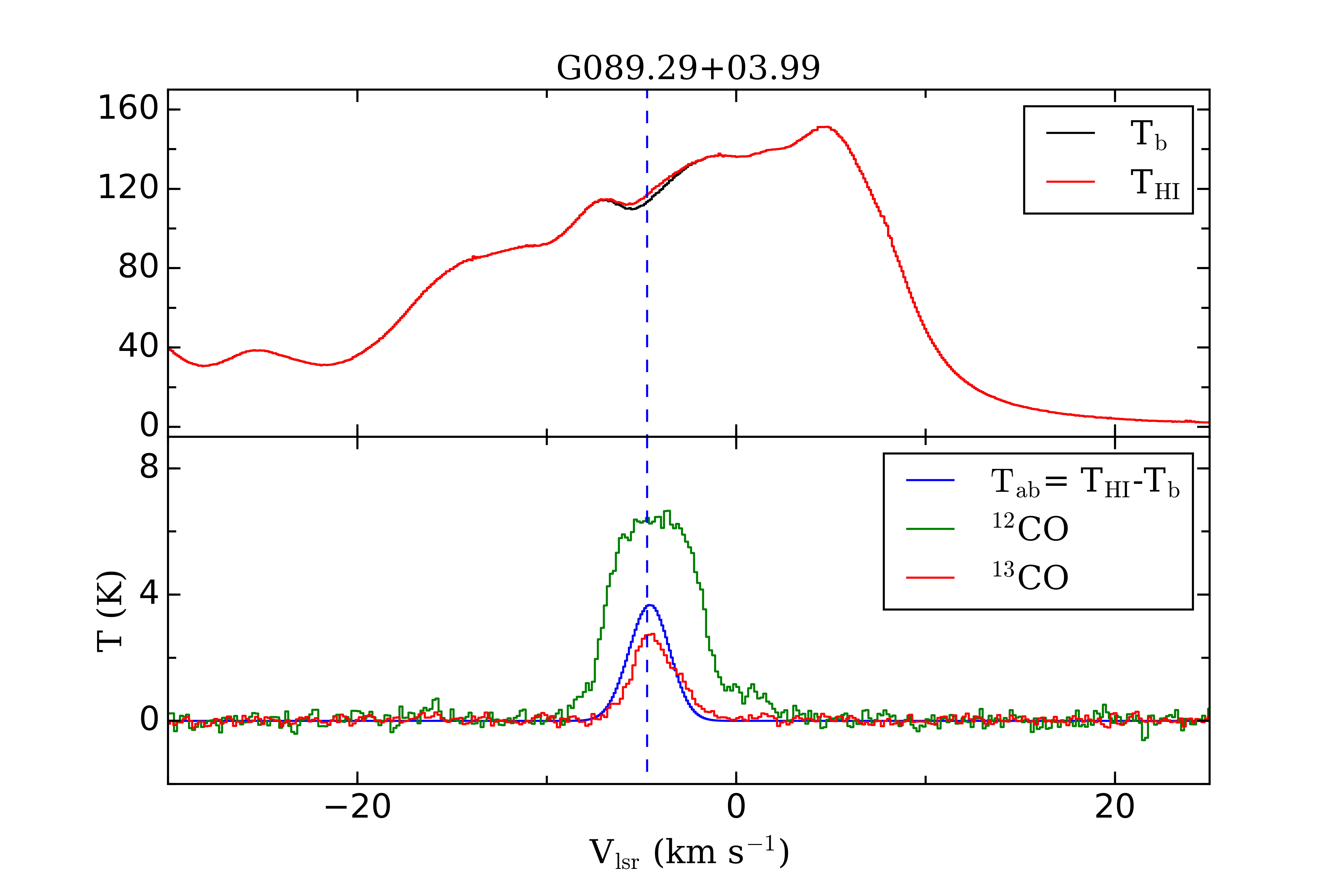}
\includegraphics[width=0.4\textwidth]{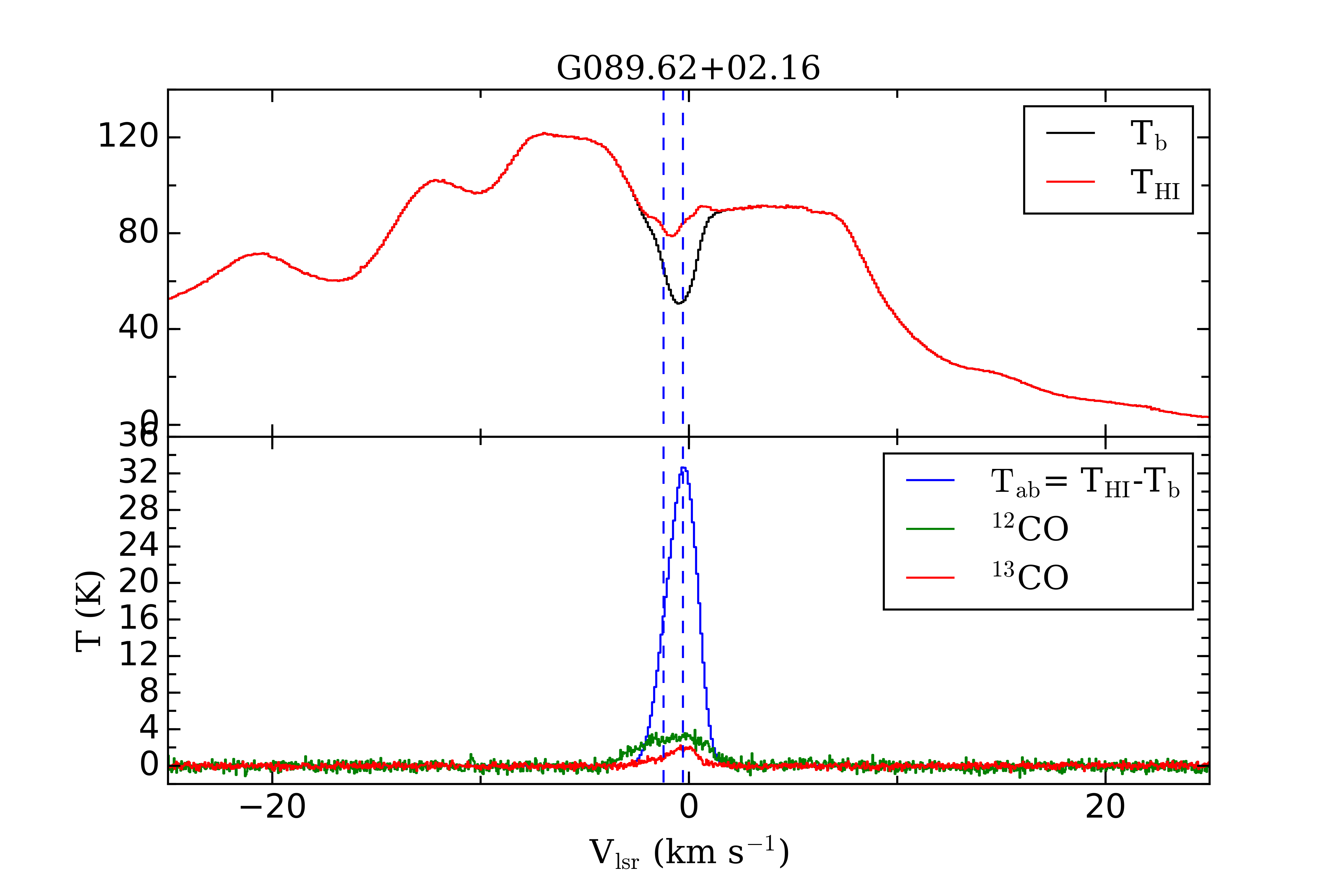}
\includegraphics[width=0.4\textwidth]{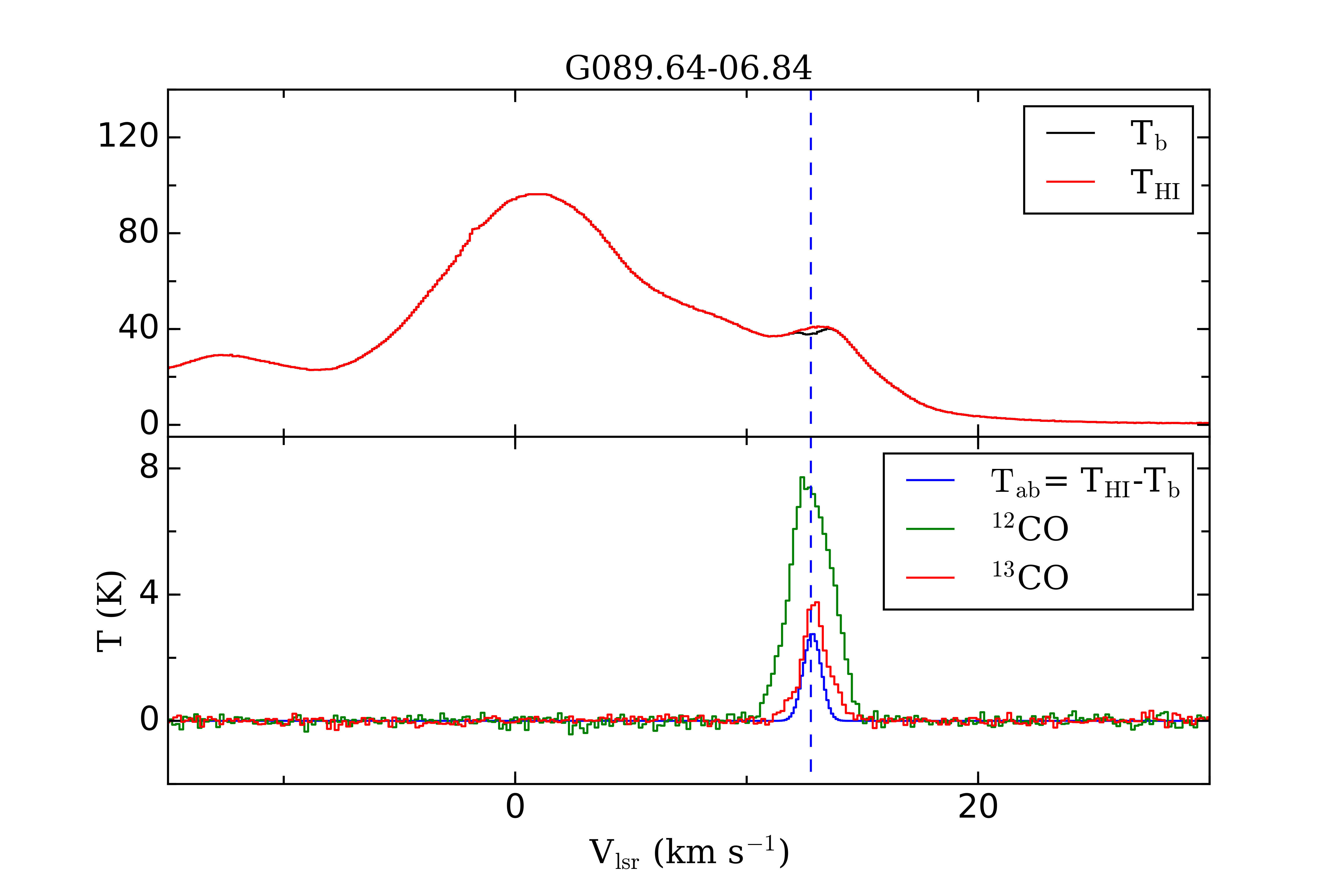}
\includegraphics[width=0.4\textwidth]{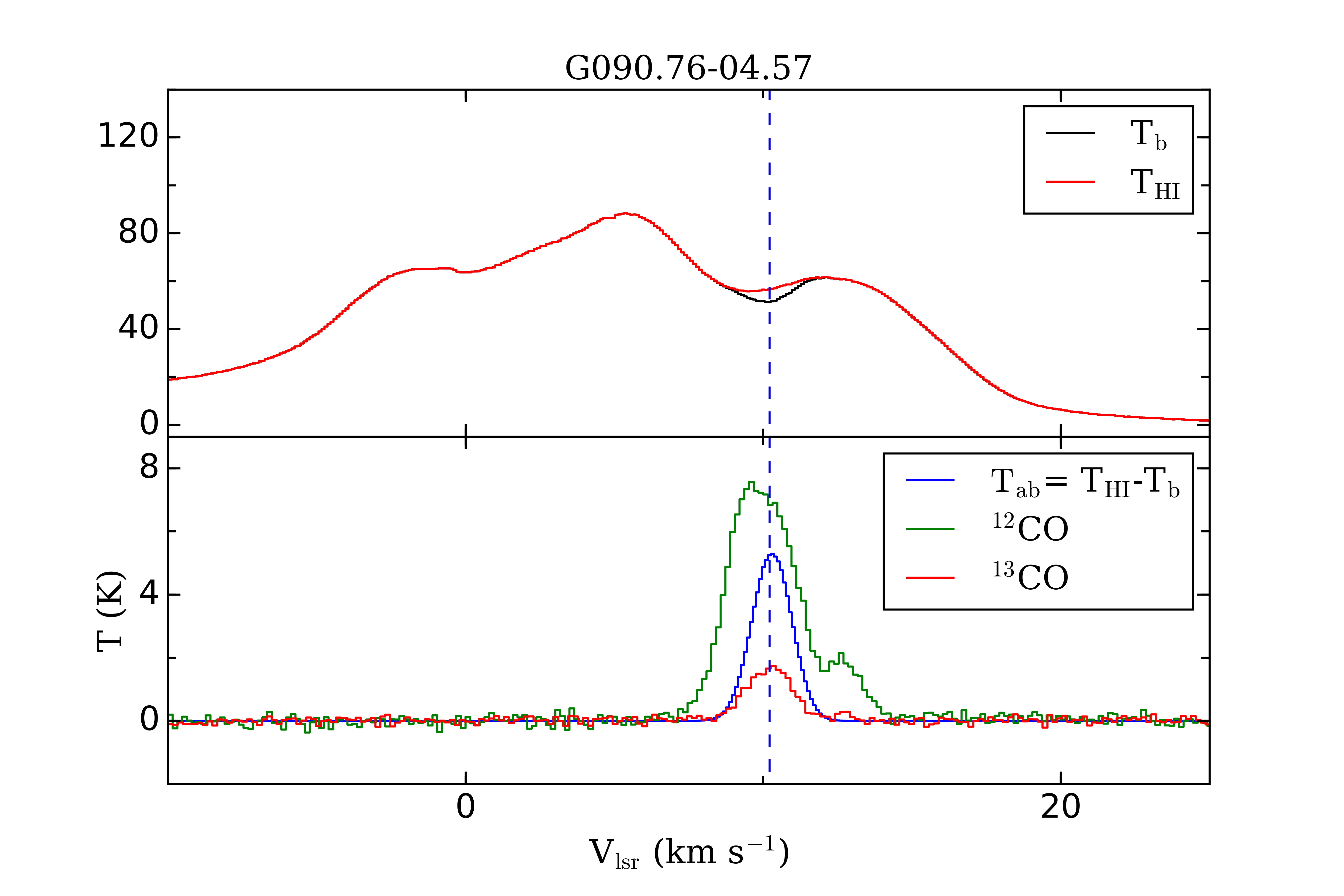}
\includegraphics[width=0.4\textwidth]{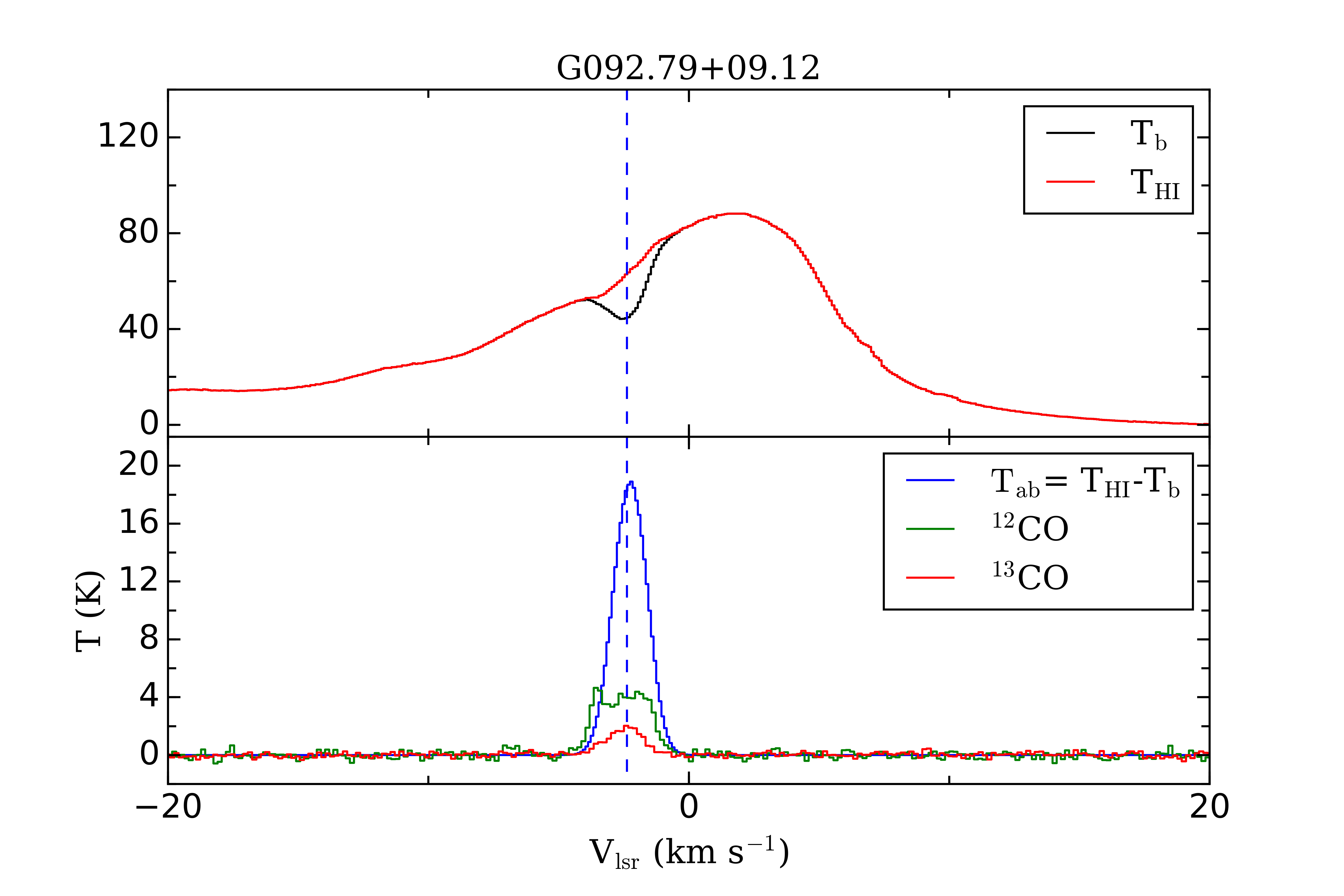}
\includegraphics[width=0.4\textwidth]{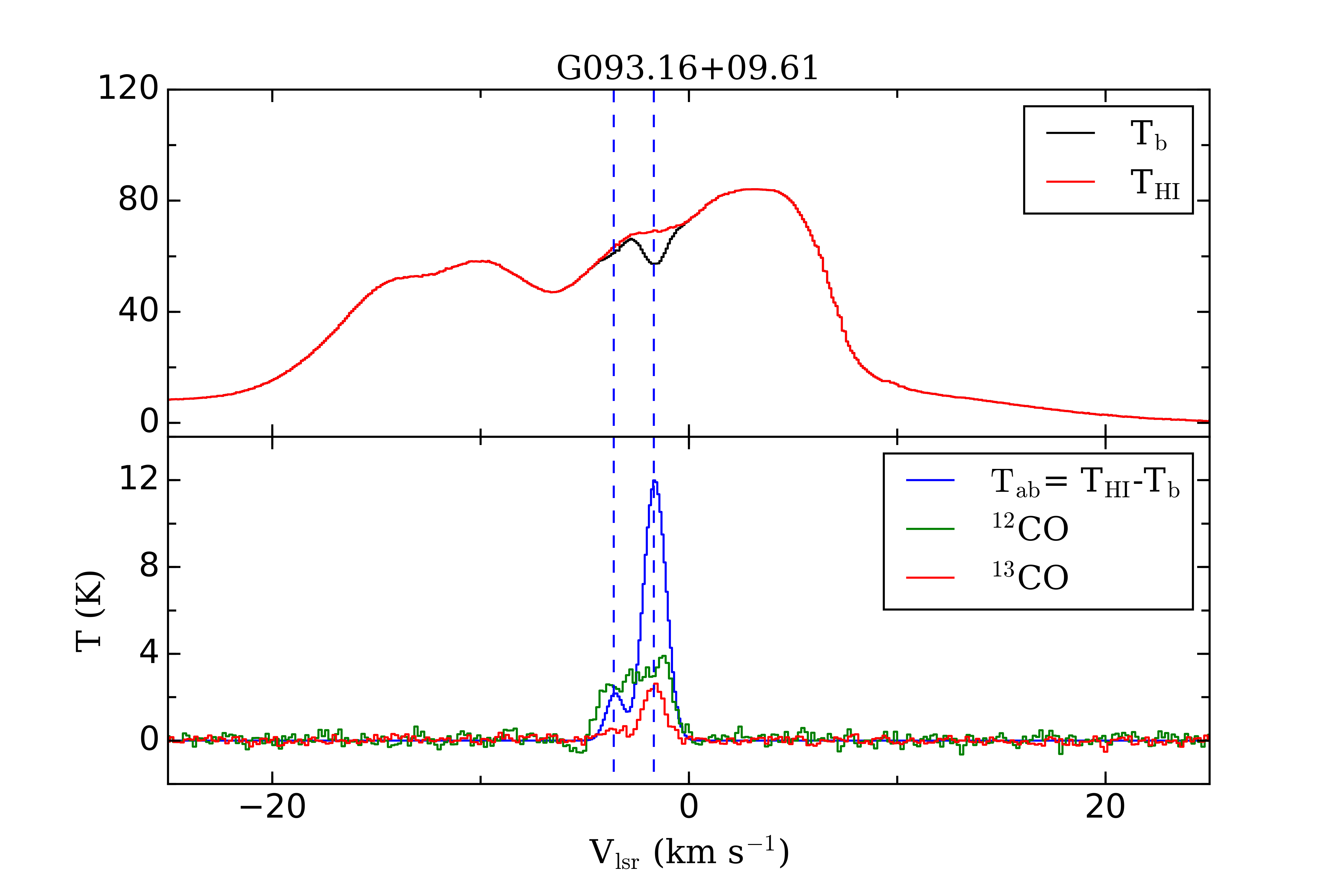}
\includegraphics[width=0.4\textwidth]{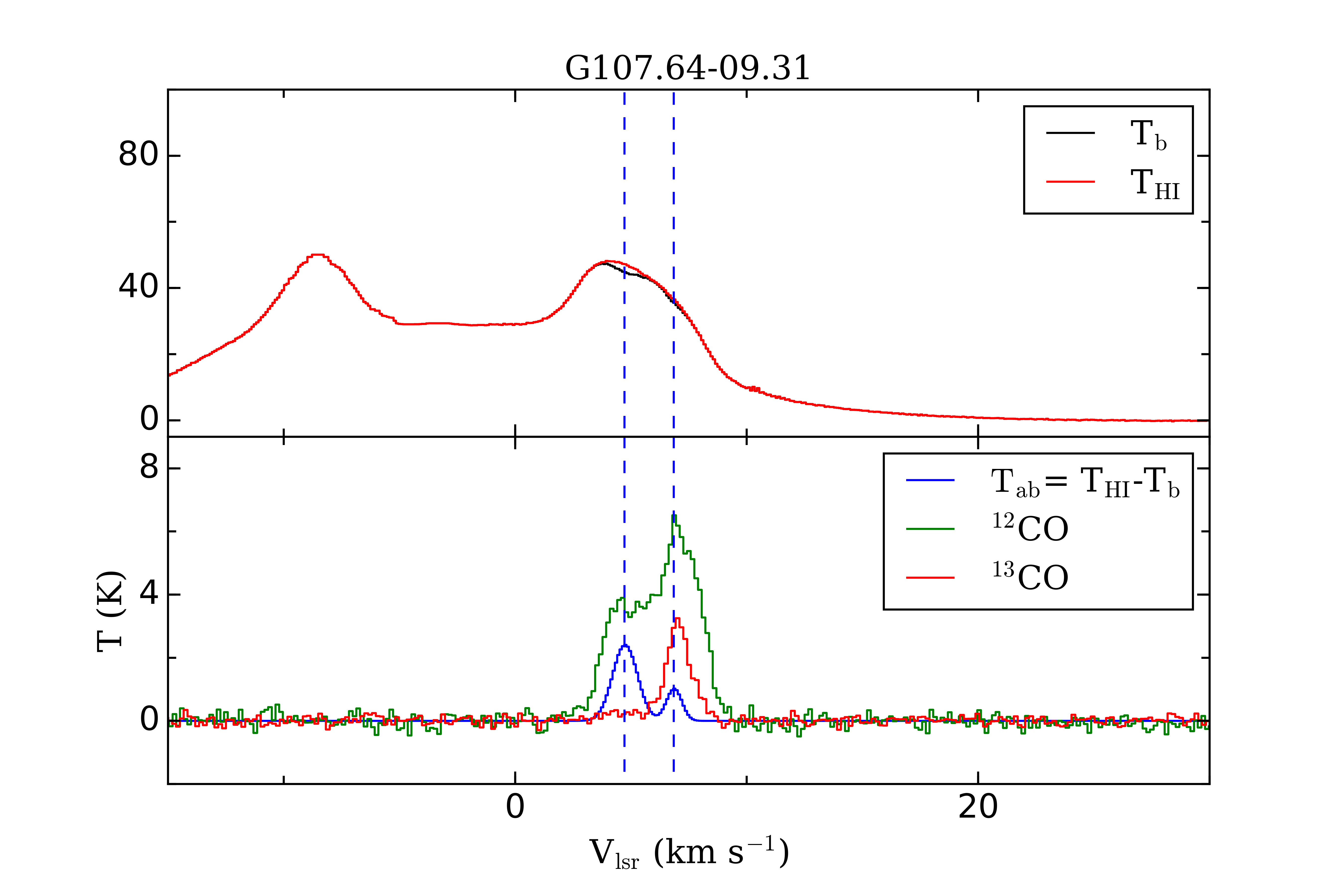}
\includegraphics[width=0.4\textwidth]{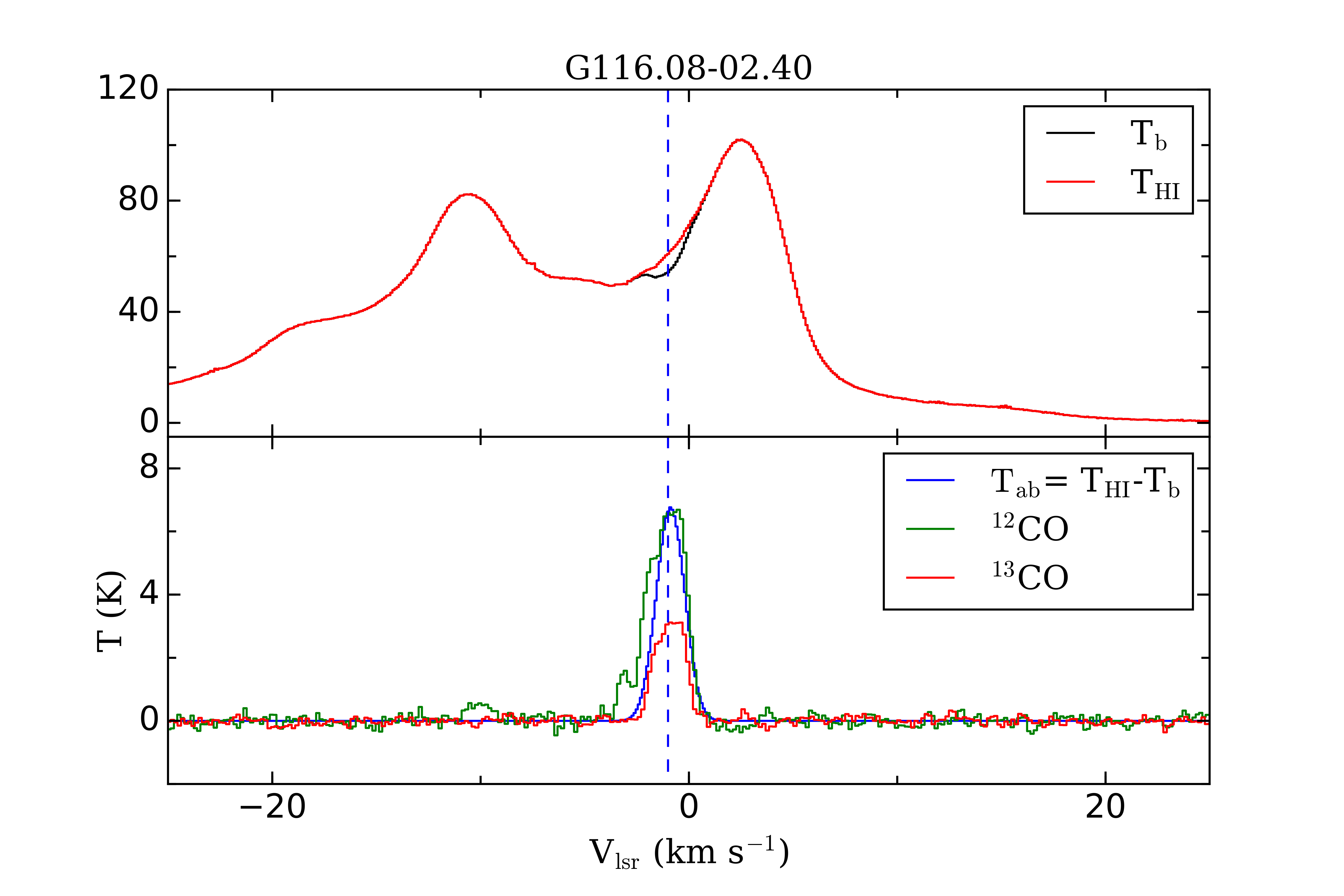}
\includegraphics[width=0.4\textwidth]{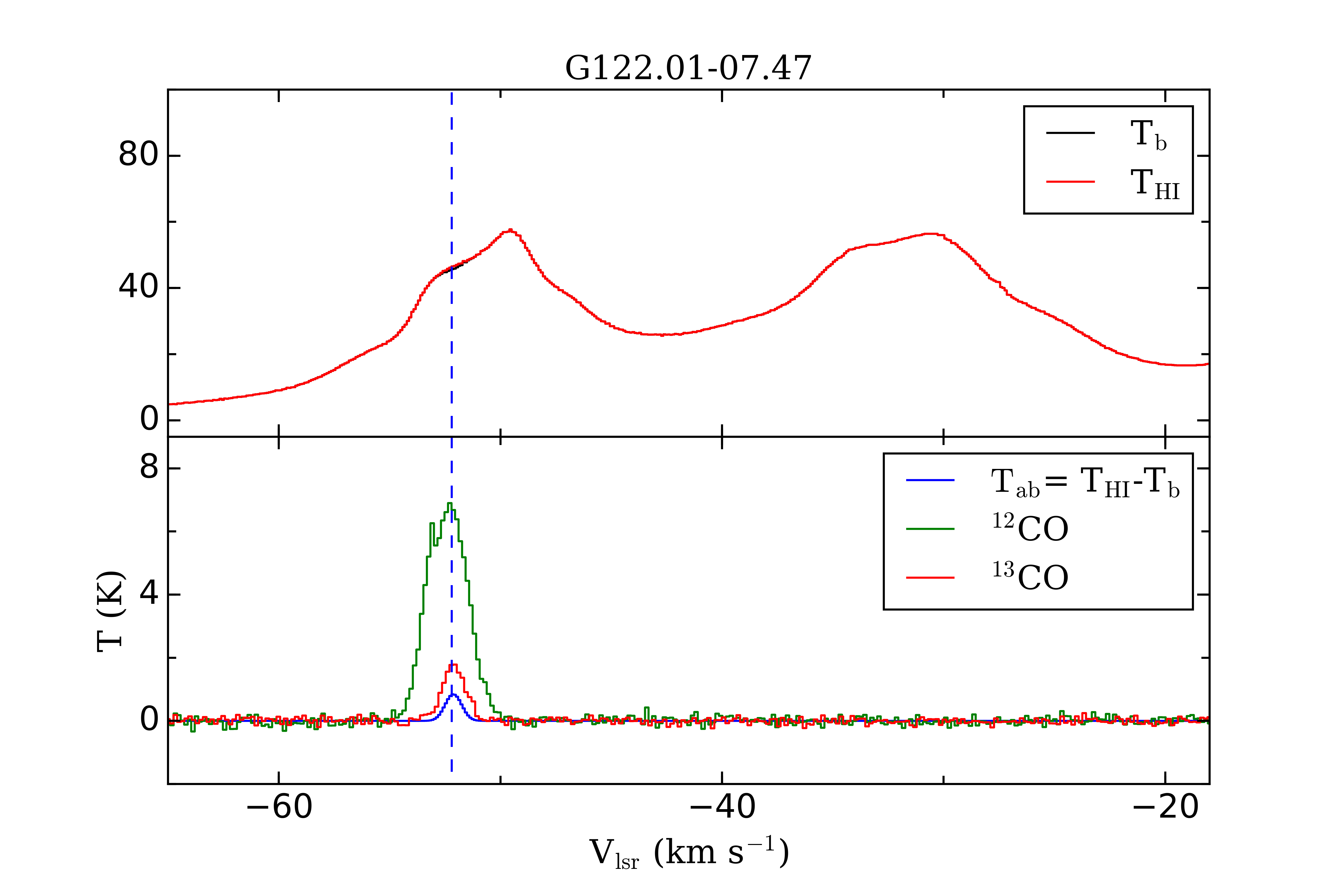}
\caption{\hi and CO spectra toward other 9 PGCCs with HINSA . Descriptions are same with that of Fig.\ref{fig:hinsa}.}
\label{fig:hinsa_13co_all}
\end{figure}

\clearpage

\section{Non-HINSA spectra toward PGCCs}
\label{sec:nonhinsa_all}
\begin{figure}[!htp]
\centering
\includegraphics[width=0.45\textwidth]{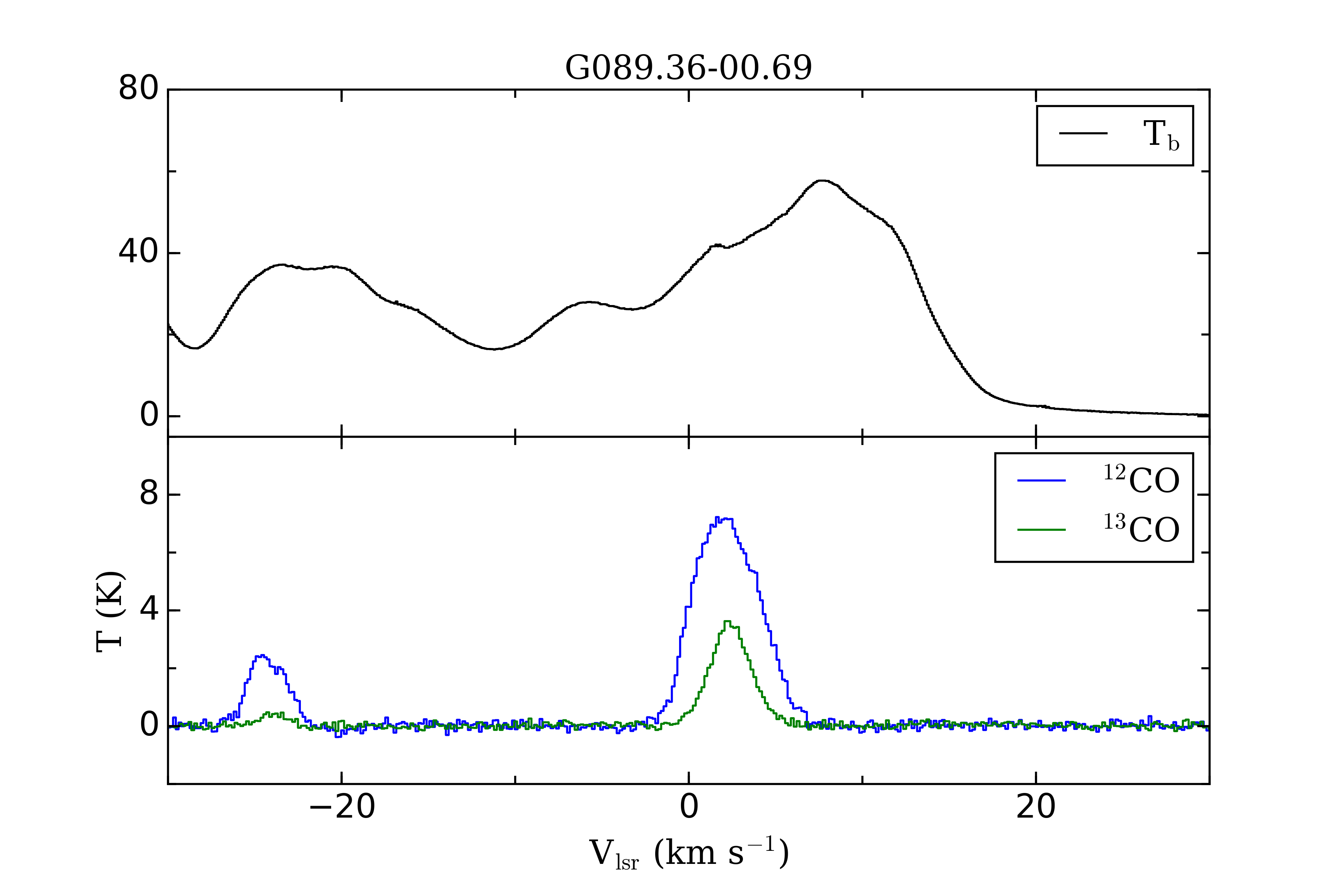}
\includegraphics[width=0.45\textwidth]{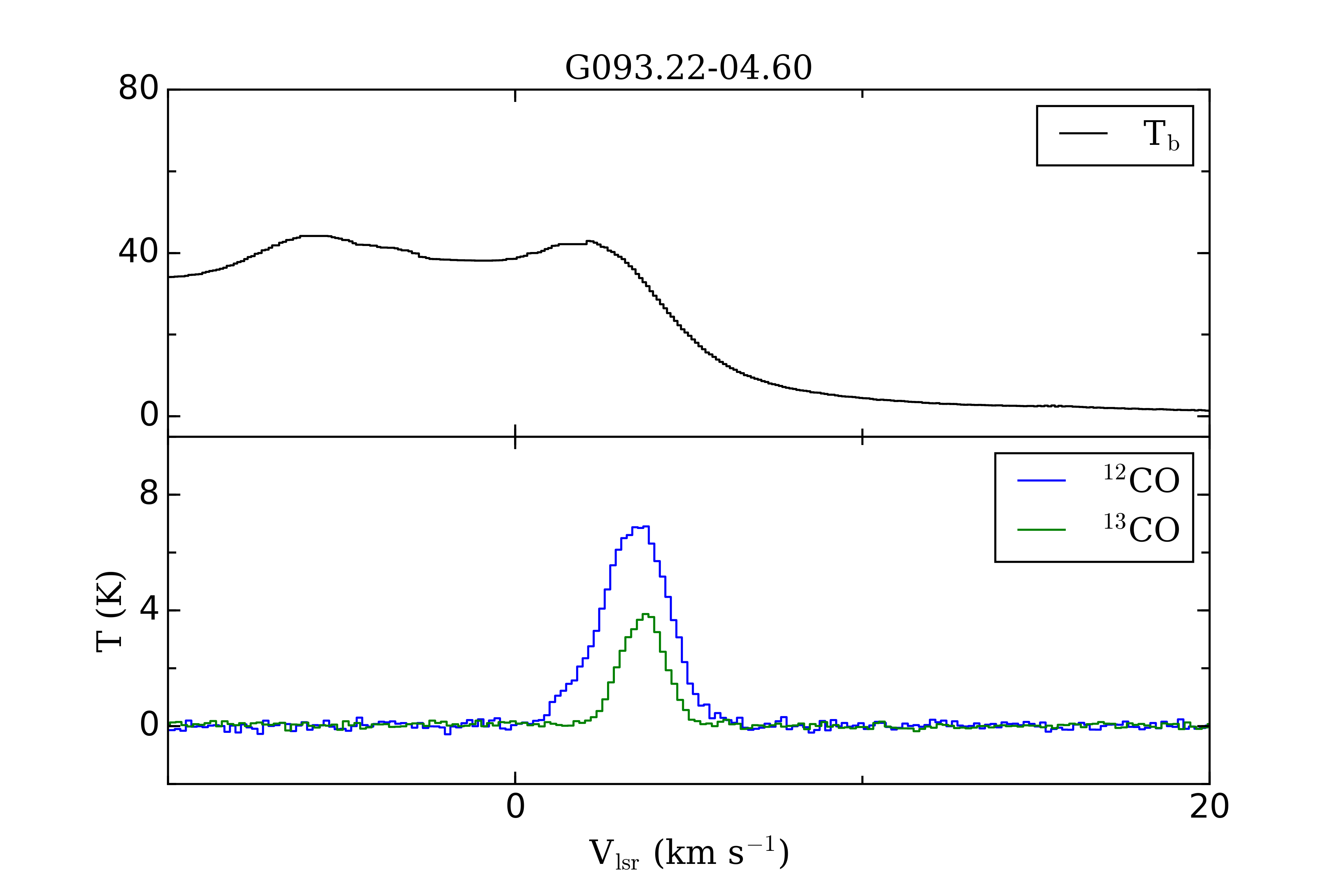}
\includegraphics[width=0.45\textwidth]{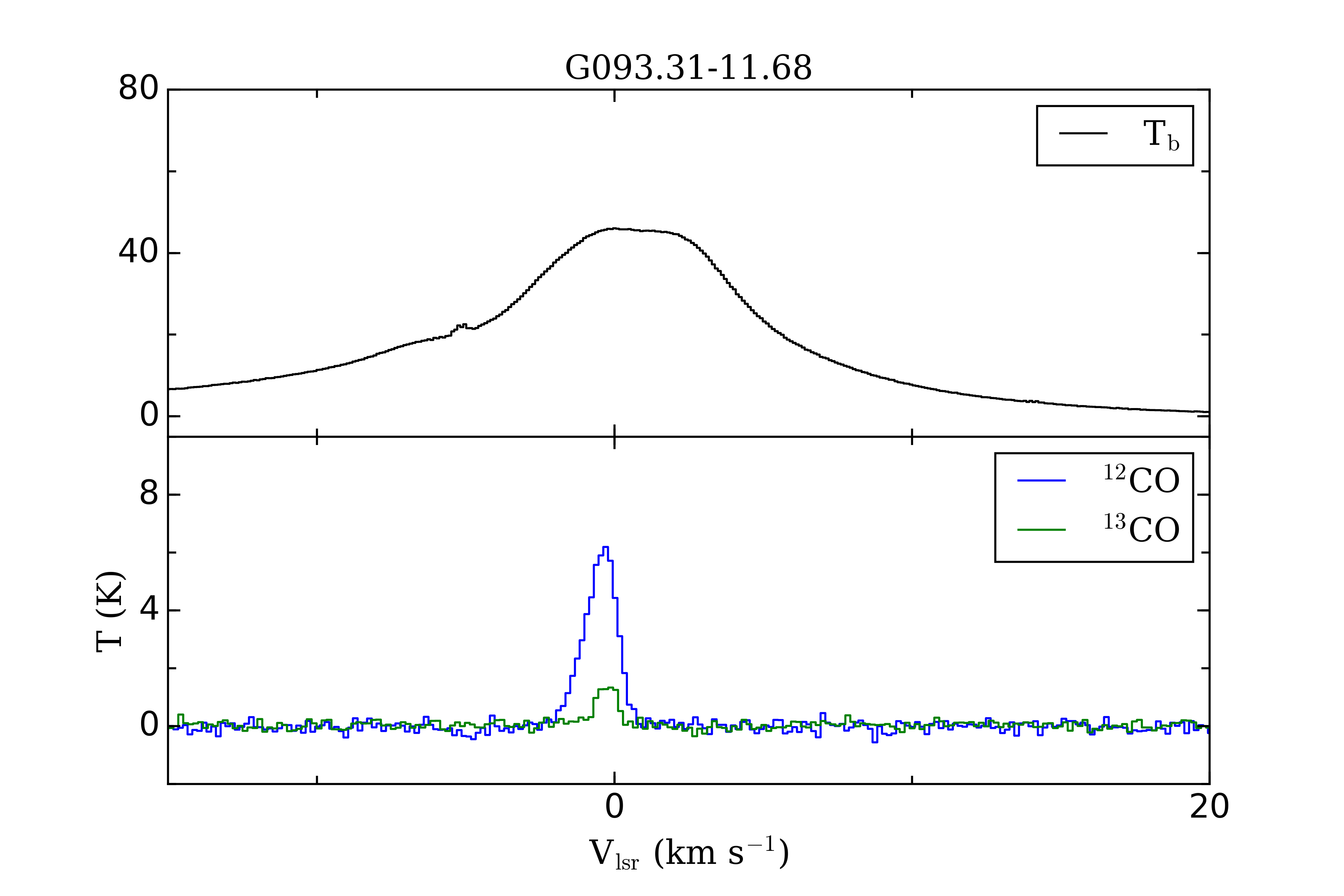}
\includegraphics[width=0.45\textwidth]{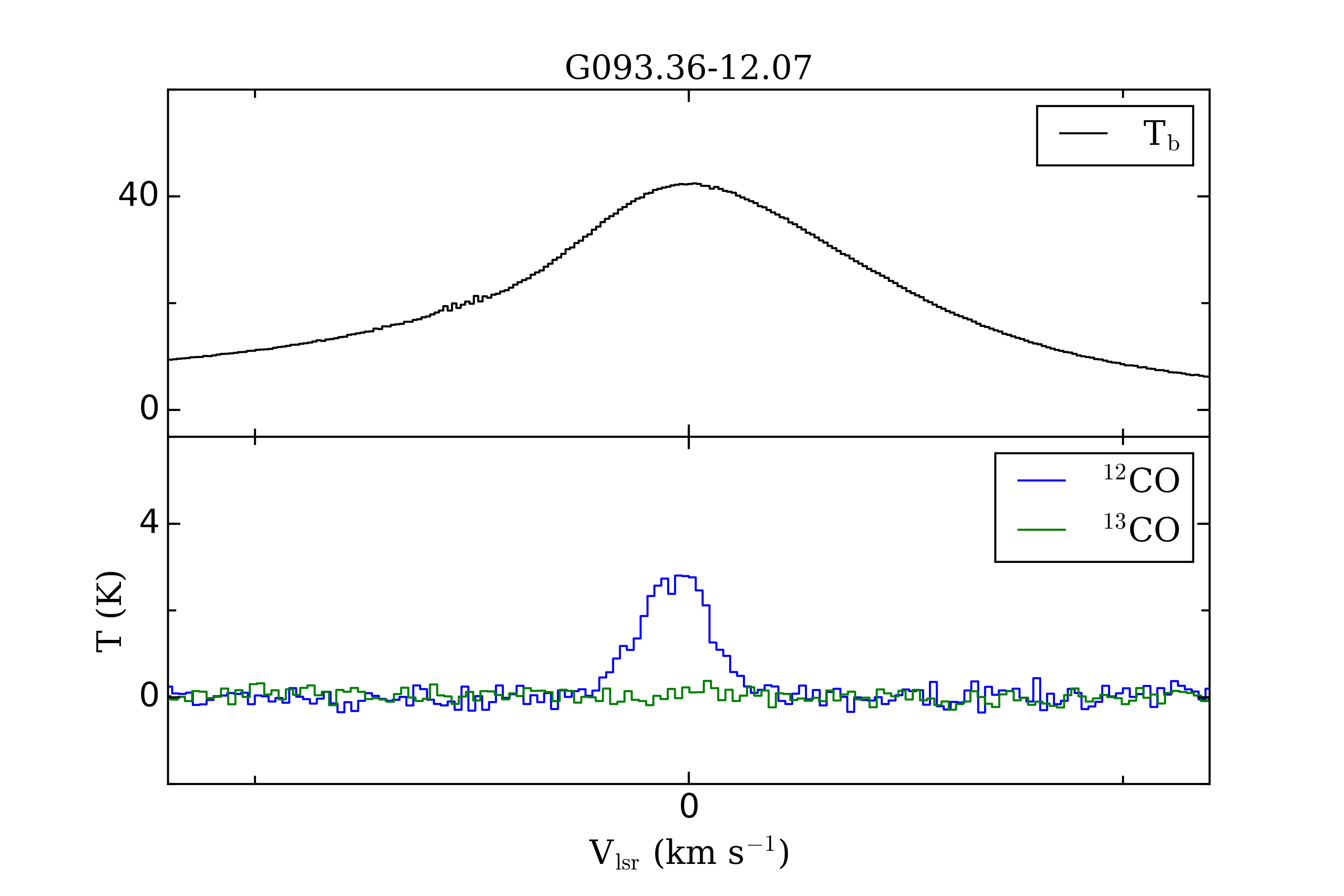}
\includegraphics[width=0.45\textwidth]{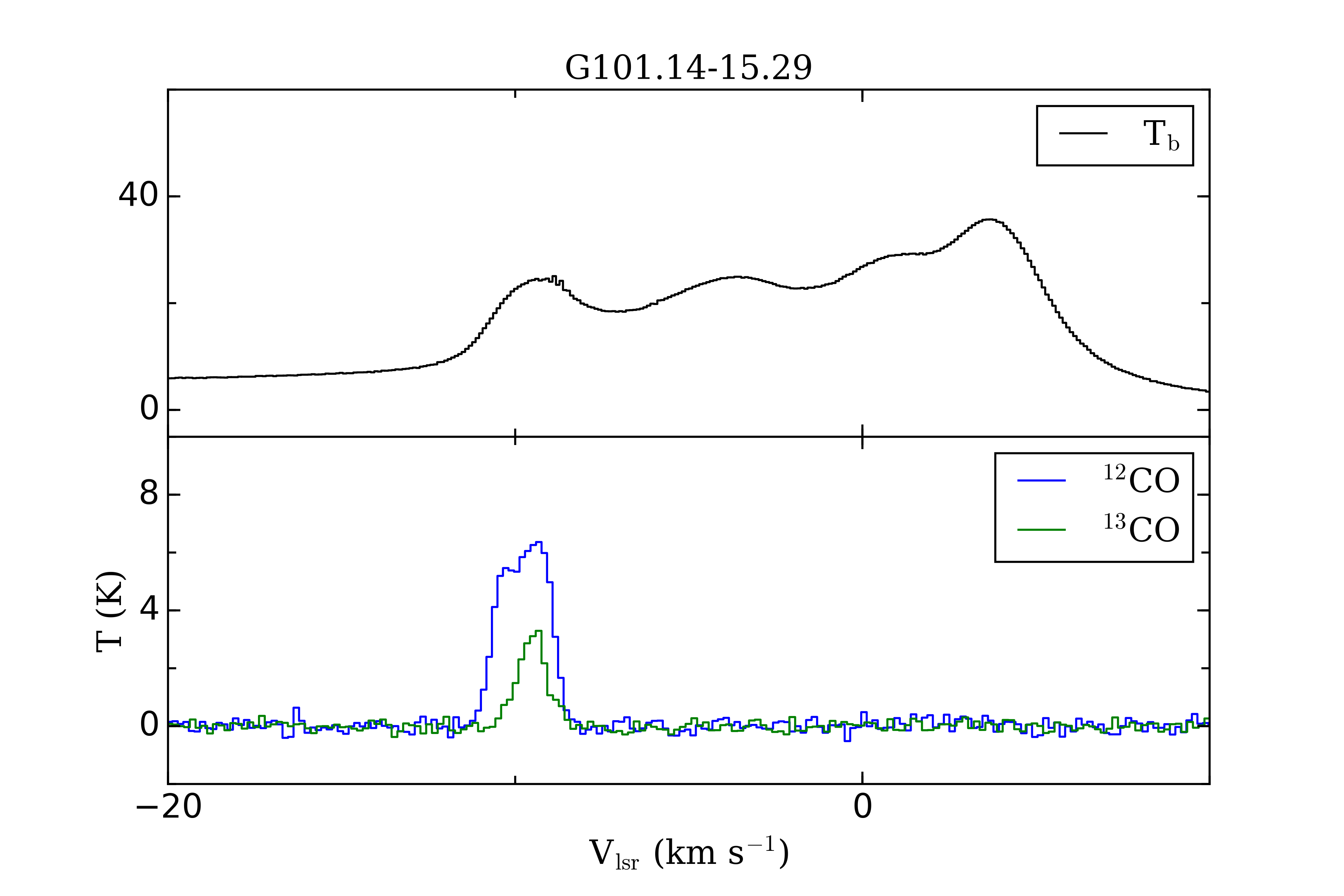}
\includegraphics[width=0.45\textwidth]{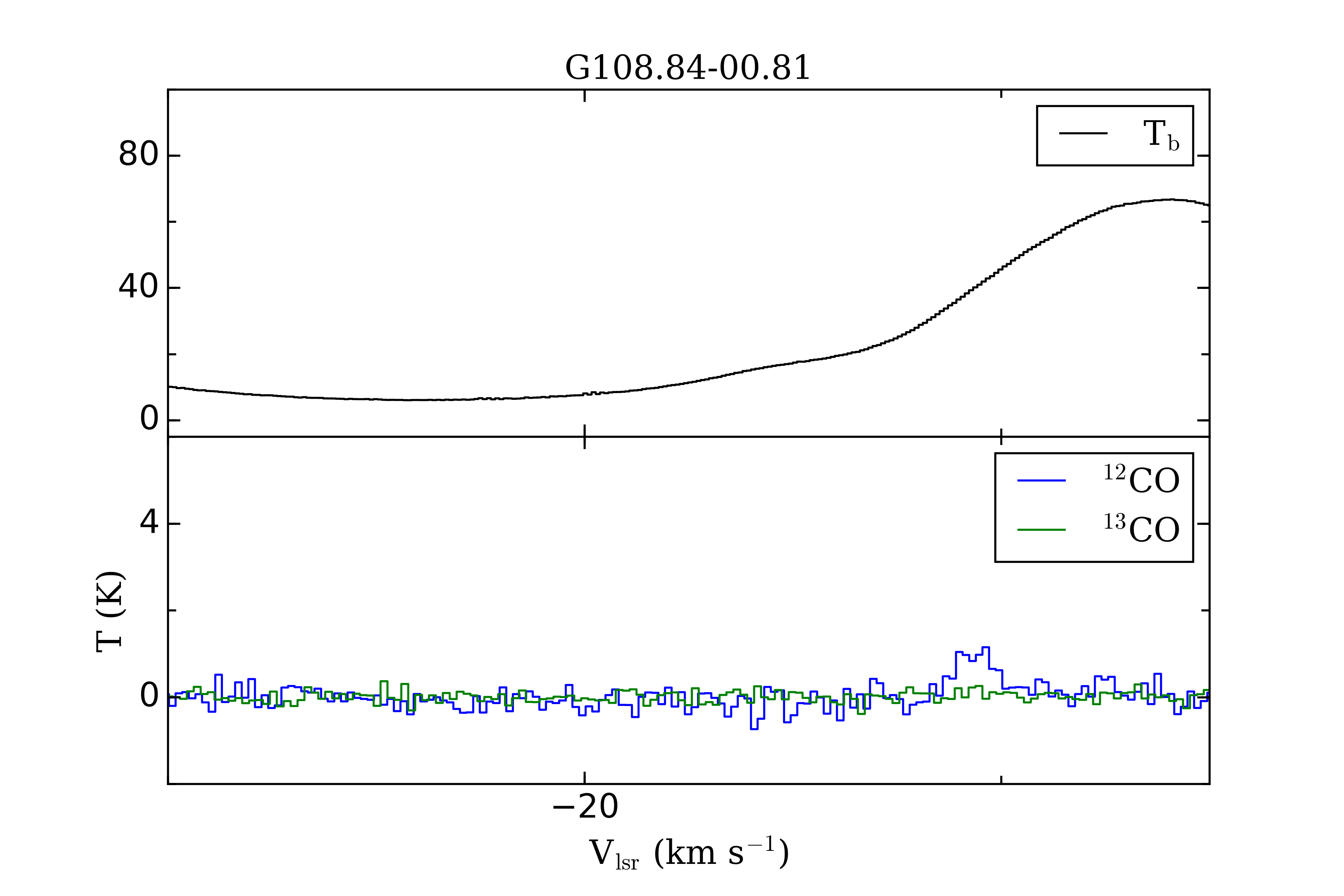}
\includegraphics[width=0.45\textwidth]{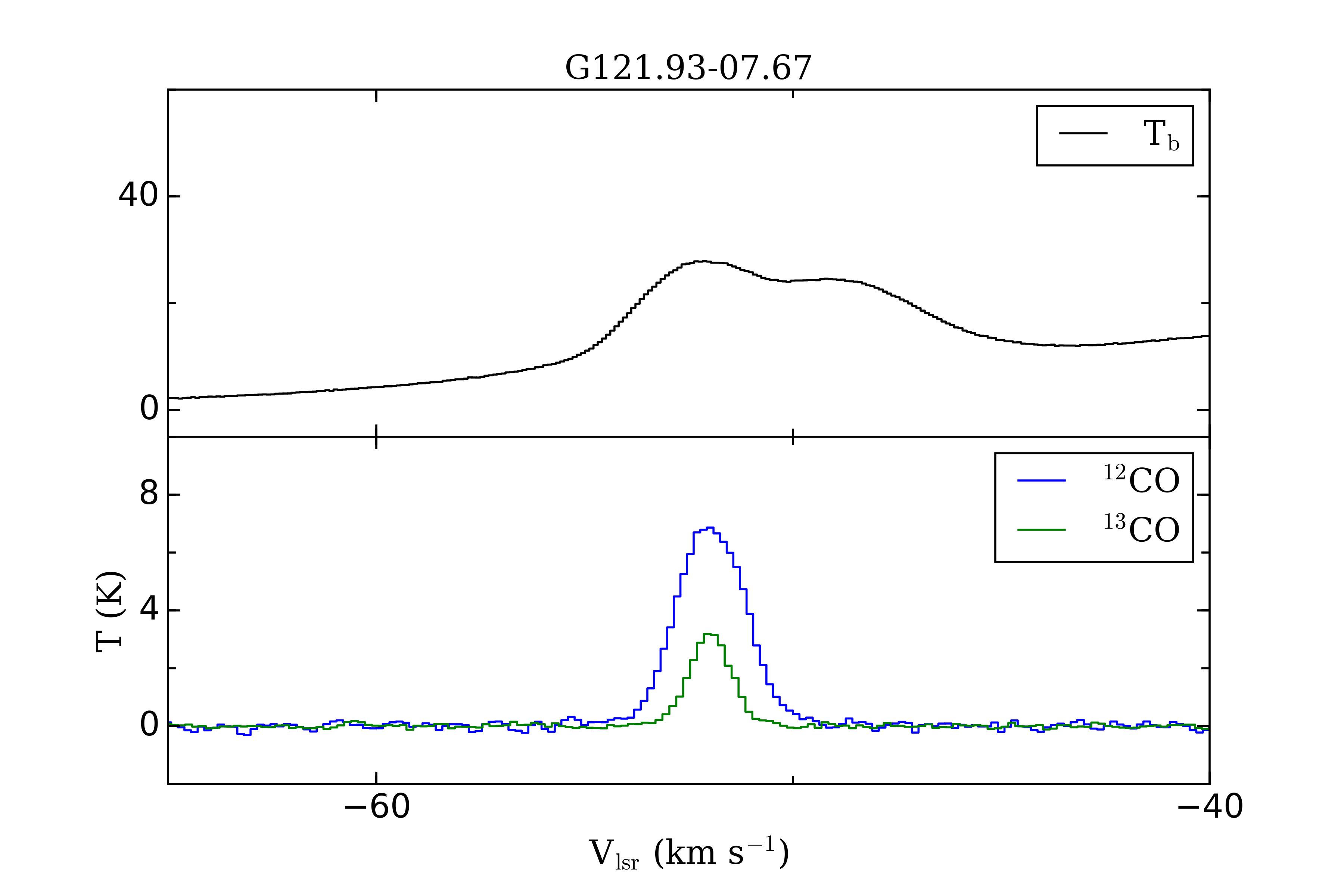}
\caption{\hi and CO spectra toward 7 PGCCs without HINSA. The observed \hi spectrum and CO spectra are shown in the top and bottom panel, respectively. }
\label{fig:13co_all}
\end{figure}

\end{document}